\documentclass[journal ]{new-aiaa}
\usepackage[utf8]{inputenc}
\usepackage{textcomp}

\usepackage{graphicx}
\usepackage{amsmath}
\usepackage[version=4]{mhchem}
\usepackage{siunitx}
\usepackage{longtable,tabularx}
\usepackage{acronym}
\usepackage{amsfonts}
\usepackage{bm}
\usepackage{algpseudocode}
\usepackage{algorithm}
\usepackage{systeme}
\usepackage{subcaption}
\usepackage{soul}
\usepackage{xcolor}
\usepackage{placeins}
\usepackage{multirow}
\usepackage{doi}
\usepackage{url}
\usepackage{fancyhdr}

\acrodef{sda}[SDA]{Space Domain Awareness}
\acrodef{rmse}[RMSE]{root mean square error}
\acrodef{epn}[EPN]{Equivalent Process Noise}
\acrodef{opm}[OPM]{Outer Probability Measure}
\acrodef{pdf}[PDF]{Probability Density Function}
\acrodef{resp}[resp.]{respectively}
\acrodef{so}[SO]{Space Object}
\acrodef{rso}[RSO]{Resident Space Object}
\acrodef{tle}[TLE]{Two-Line Element}
\acrodef{smc}[SMC]{Sequential Monte Carlo}
\acrodef{gmm}[GMM]{Gaussian Mixture Model}
\acrodef{iod}[IOD]{Initial Orbit Determination}
\acrodef{mm}[MM]{Multiple Model}
\acrodef{gpu}[GPU]{Graphic Processing Unit}
\acrodef{mh}[MH]{Multiple Hypothesis}
\acrodef{mht}[MHT]{Multiple Hypothesis Tracker}
\acrodef{imm}[IMM]{Interacting Multiple Model}
\acrodef{vs}[VS]{Variable Structure}
\acrodef{fs}[FS]{Fixed Structure}
\acrodef{mf}[MF]{Multi-Fidelity}
\acrodef{lf}[LF]{Low-Fidelity}
\acrodef{hf}[HF]{High-Fidelity}
\acrodef{jms}[JMS]{Jump-Markov System}
\acrodef{ukf}[UKF]{Unscented Kalman Filter}
\acrodef{ekf}[EKF]{Extended Kalman Filter}

\title{A Bayesian Approach to Low-Thrust Maneuvering Spacecraft Tracking}

\author{Enrico M. Zucchelli \footnote{PhD Candidate, Aerospace Engineering and Engineering Mechanics, 2617 Wichita St.; email: enricomarco@utexas.edu} and Brandon A. Jones\footnote{Associate Professor, Aerospace Engineering and Engineering Mechanics, 2617 Wichita St., Associate Fellow AIAA\\Manuscript published in the AIAA \textit{Journal of Guidance, Control, and Dynamics}. DOI: {10.2514/1.G007849}}}
\affil{The University of Texas at Austin, Austin, TX, 78705}

\begin{document}

\maketitle
\thispagestyle{fancy} 

\begin{abstract}
Bayesian estimation with an explicit transitional prior is required for a tracking algorithm to be embedded in most multi-target tracking frameworks.
This paper describes a novel approach capable of tracking maneuvering spacecraft with an explicit transitional prior and in a Bayesian framework, with fewer than two observations passes~per day.
The algorithm samples thrust profiles according to a multivariate Laplace distribution.
It is shown that multivariate Laplace distributions are particularly suited to track maneuvering spacecraft, leading to a log~probability function that is almost linear with the thrust.
Principles from rare event simulation theory are used to propagate the tails of the distribution. Fast propagation is enabled by multi-fidelity methods.
Because of the diffuse transitional prior, a novel $k$-nearest neighbor-based ensemble Gaussian mixture filter is developed and used.
The method allows Bayesian tracking of maneuvering spacecraft for several scenarios with fewer than two measurement passes per day, and with a mismatch between the true and expected thrust magnitude of up to a factor of 200. The validity domain and statistical significance of the method are shown by simulation through several Monte Carlo trials in different scenarios and with different filter settings.
\end{abstract}

\section*{Nomenclature}

{\renewcommand\arraystretch{1.0}
\noindent\begin{longtable*}{@{}l @{\quad=\quad} l@{}}
$a$ & orbital semi-major axis [km]\\
$C_D$ & drag coefficient [-]\\
$C_R$ & solar radiation pressure coefficient [-]\\
$e$ & orbital eccentricity [-]\\
$F$ & state transition matrix\\
$\bm{f}\left(\cdot\right)$ & state transition function \\
$G$ & Gramian matrix \\
$G(\cdot)$ & gating function \\
$i$ & orbital inclination [rad] \\
$\bm{h}\left(\cdot\right)$ & measurement function \\
$H$ & measurement Jacobian \\
$I$ & indicator function \\
$\mathbb{I}$ & identity matrix \\
$K(\cdot)$ & kernel function \\
$N$ & number of samples \\
$p\left(\cdot\right)$ & probability distribution \\
$P$ & covariance matrix\\
$q\left(\cdot\right)$ & importance sampling distribution \\
$Q$ & process noise matrix\\
$\bm{r}$ &   position [km]\\
$R$ & measurement noise matrix\\
$R_e$ & planet equatorial radius [km] \\
$r$ & range [km]\\
$\bm{t}$ & thrust profile [km/s\textsuperscript{2}] \\
$t$ & time [s]\\
$w$ & particle/component weight \\
$\bm{v}$ &    velocity [km/s]\\
$\bm{y}$ & measurement vector \\
$\alpha$ & right ascension [rad]\\
$\bm{\gamma}$ & important sample for the stochastic collocation method\\
$\delta$ & declination [rad]\\
$\delta(\cdot)$ & Dirac delta function\\
$\epsilon$ & measurement noise \\
$\lambda$ & Laplace distribution inverse scale parameter \\
$\Lambda(\cdot)$ & likelihood function \\
\textmu   & gravitational parameter [km\textsuperscript{3}/s\textsuperscript{2}] \\
$\nu$ & true anomaly [rad]\\
$\sigma$ & standard deviation \\
$\tau$ & exponential random variable \\
$\bm{\chi}$ & augmented state \\
$\Omega$ & right ascension of ascending node [rad]\\
$\omega$ & argument of periapsis [rad]\\
$\cdot_p$ & pertaining to the measurement pass \\
$\cdot_t$ & pertaining to the trajectory\\
$\cdot^H$ & high-fidelity model\\
$\cdot^L$ & low-fidelity model
\end{longtable*}}

\section{Introduction}
Satellite maneuver capabilities are ever evolving, and not all operators are willing to share data about the paths of their spacecraft. Tracking is a necessary tool for space traffic management and collision avoidance, but unexpected maneuvers may lead to increased uncertainty and even loss of custody. The problem is exacerbated by the low frequency of observations passes, and by the variety of modality of maneuvers, ranging from high-thrust and impulsive, to low-thrust and with continuously varying magnitude and direction. A satellite's identity cannot be determined based on a measurement via traditional instruments (\textit{e.g.}, electro-optical and radar sensors), and thus the association between a measurement and a spacecraft has to rely on prior information. It is the propagation of an explicit initial prior into a transitional prior that allows proper data association, required for most multi-target trackers such as the classical multiple hypothesis tracker~\cite{reid_1979} or the generalized labeled multi Bernoulli (GLMB) filter~\cite{vo_2017,ravago2021risk}. While the requirement of an explicit transitional prior can be circumvented~\cite{holzinger_2012,escribano2022automatic}, the absence of it generally leads to non-Bayesian, heuristics-based filters. The goal of this work is to develop a method that can efficiently track low-thrust maneuvering spacecraft when the observation passes are sparse and the maneuver is unknown, while at the same time being compatible with Bayesian multi-target tracking algorithms.

Maneuvering target tracking is a widely researched field~\cite{li_2003,li2001survey,li2001bsurvey,li2002survey}; however, little literature is available on Bayesian maneuvering target tracking for space objects. Challenges specific to the low-thrust maneuvering satellite problem are due to: a) infinite dimensionality of the maneuver profile, b) data sparsity, c) nonlinearity of the problem, and d) potentially large mismatches between physics models and actual behavior of the target when a maneuver is performed. A large portion of the literature on maneuvering target tracking relies on the Kalman filter~\cite{kalmanfilter} and on the extended Kalman filter~(EKF)~\cite{barshalomnonlinear}.
The simplest way to track maneuvering targets is by equivalent process noise, in which the process noise of a Kalman filter is inflated depending on how large the difference between actual and expected measurement is~\cite{gholson1977maneuvering,efe1998maneuvering}. While fast and easy to implement, this approach is only successful when the maneuvers are small and when the frequency of observation passes is high, since large state uncertainties quickly become non-Gaussian. Moreover, the method requires that measurement and target are already associated, because the inflation of the target uncertainty is a function of the measurement residuals. Variable dimension Kalman filters estimate maneuvers together with the state of the target~\cite{bar1982variable,goff2015orbit}. To allow for observability, a constant acceleration through multiple measurements has to be assumed, which is not generally applicable when working with sparse data and low-thrust propelled space objects. On the other hand, it is possible to track spacecraft after single impulsive maneuvers followed by long duration coasting arcs~\cite{pastor2022satellite,porcelli2022}. Optimal control-based approaches, such as those detailed in Refs.~\cite{lubey_2014} or \cite{pirovano2022detection}, allow for continuously varying maneuvers between two measurements, but also require the data association to be known, and assume optimality.
Interacting Multiple Model filters~\cite{li_2005} are another popular methodology to track non-cooperative objects; this approach is however only optimal when the number of models is finite, which is not the case for a spacecraft. While the technology has been applied to maneuvering spacecraft by merging different maneuvers together~\cite{goff_2015,zucchelli2020tracking}, the finite number of models still introduces limitations to the applicability of the method.

A less common approach to maneuvering target tracking is to use maneuver distributions that are not sub-Gaussian. For the goal of this paper, a sub-Gaussian distribution can informally be described as a distribution with tails that decay as fast as, or faster than, Gaussian tails. In such a framework, the maneuver is seen as process noise, like in equivalent process noise Kalman filters, but its distribution has tails that decay slower than those of a Gaussian distribution. At the same time, the distribution of the process noise is not adapted to the measurement. Filters with non sub-Gaussian, or super-Gaussian, process noise can keep custody of a target better than corresponding Gaussian filters for large deviations in the dynamics. Roth et al.~\cite{roth2013student} develop a closed form linearized filter with Student's $t$ process and measurement noises; Huang et al.~\cite{huang2017robust} employ variational Bayesian methods for an optimization-based filter with multivariate Laplace process noise. With rare exceptions, (\textit{e.g.}, Refs.~\cite{ikoma2002tracking,zucchelli2023gif}), the use of non sub-Gaussian priors for the maneuvers is limited to linear or linearized settings. Representing nonlinear transformations with high accuracy for the super-Gaussian tails can be a challenging task, as they can be orders of magnitude larger than the tails of Gaussian distributions.

In this work, the utility of super-Gaussian process noise distributions, and specifically multivariate Laplace distributions, is extended to nonlinear problems by exploiting concepts from rare event simulation techniques~\cite{rubino2009rare,beck2015rare} and $k$-nearest neighbor kernel density estimation. The computational burden caused by the large amount of sample propagation required is mitigated by the use of multi-fidelity methods~\cite{peherstorfer_2018}. 

The main contribution of this paper is to provide an approach to Bayesian maneuvering target tracking that is capable of tracking low-thrust maneuvering spacecraft when observation passes are very sparse and the thrust profile and magnitude are unknown. To the best of the authors' knowledge, this is the first successful attempt to do so. Bayesian estimation is necessary to multi-target tracking; as satellite tracking is ultimately a multi-target tracking problem, being compatible with multi-target tracking algorithms is a requirement for operational applications of the method.
There are six additional contributions in this paper. First, it is shown how to apply multi-fidelity methods to the propagation of a maneuvering satellite, where the dimensionality of the problem increases greatly because of the maneuvers. Second, it is argued and shown that a multivariate Laplace distribution is especially suited to represent the process noise for maneuvering spacecraft, as it is close to being linearly exponential in the thrust, like the rocket equation. Third, it is proven that, when performing Bayesian estimation with multivariate Laplace priors and Gaussian measurements, rare event simulation and importance sampling greatly reduce the variance and bias of the estimates when large deviations, such as they may be observed in maneuvering target tracking, occur. Fourth, a novel ensemble Gaussian mixture filter algorithm, based on $k$-nearest neighbors, is proposed and developed. Fifth, a nonlinear filter that is robust to large deviations thanks to multivariate Laplace process noise is derived. The filter has an explicit transitional prior that is independent of the measurement, making it compatible with Bayesian multi-target tracking frameworks, and exploits specific synergy between the rare event simulation technique and the $k$-nearest neighbor ensemble Gaussian mixture filter. Sixth, the filter is tested over several challenging scenarios, with large differences between the assumed maneuver magnitude and the actual maneuver magnitude, and with very sparse observation passes.

The paper proceeds as follows. Section~\ref{sec:MF} describes the multi-fidelity stochastic collocation method, and how it is adapted to the maneuvering target tracking problem. Section~\ref{sec:thrustdistr} describes the distribution of the thrust profile, and Sec.~\ref{sec:rareevent} introduces a method to efficiently sample from Laplace distributions. In Sec.~\ref{sec:nnengmf}, a novel ensemble Gaussian mixture filter is described. Sec.~\ref{sec:filtersummary} summarizes the filter, and Section~\ref{sec:results} reports the results obtained by simulation. Finally, conclusions are drawn in Sec.~\ref{sec:conclusions}.

\section{Multi-Fidelity Stochastic Collocation}
\label{sec:MF}
The most computationally intensive process in orbit determination is the propagation of the state distribution forward in time. To reduce runtime, the proposed filter uses multi-fidelity methods~\cite{peherstorfer_2018}, which are particularly useful when a large number of samples is needed, like in a particle filter~\cite{liu1998} or in a \ac{gmm} filter~\cite{sorenson_1971}. Even if spacecraft tracking is generally performed on the ground, where there are few constraints on computational cost, a speed-up allows one to use more samples, therefore achieving higher accuracy for same computational costs.
Multi-fidelity methods combine models of varying fidelity and produce trade-offs between existing models. High fidelity models are expensive and accurate; low-fidelity models offer shorter runtime but incur larger errors. This work employs multi-fidelity stochastic collocation~\cite{narayan_2014}, a method that opportunely selects a few samples to be propagated with the higher accuracy model; the outputs from those samples are then used as data to generate a linear mapping that corrects the remaining low-fidelity samples. The corrected samples, or surrogates, provide a local approximation that is almost as accurate as the high-fidelity samples, at a fraction of the computational cost.
The stochastic collocation method, which is one approach to uncertainty quantification via surrogate modeling, is chosen over other multi-fidelity methods because it has already been proven and tested in astrodynamic problems~\cite{jones_2019a}, where it offered speed-ups by a factor between 50 and 100.
The multi-fidelity stochastic collocation method has additionally already been utilized for orbit determination~\cite{zucchelli2021multi} and for uncertainty quantification in cislunar space~\cite{wolf2022multi}; in recent work, computational speed has been further increased using Vinti theory~\cite{vinti1959new,wolf2023}.

\subsection{High-Level Description of multi-fidelity stochastic collocation}
Let $\bm{x}_t=\bm{f}(\bm{x}_0): \bm{x}_0\rightarrow \mathbb{R}^m$ be a function that is evaluated for the random input $\bm{x}_0\in \mathbb{R}^d$. Both a high- and a low-fidelity model are available to evaluate the function, referred to by superscripts $\left(\cdot\right)^H$ and $\left(\cdot\right)^L$, respectively. The subscript $(\cdot)_0$ stands for the initial time, and the subscript $(\cdot)_t$ stands for the state evaluated at discrete times along the trajectory. 
The bifidelity stochastic collocation method consists of three steps:
\begin{enumerate}
    \item select a finite set $X_0=\left[\bm{x}_0^1,\,\cdots\,,\bm{x}_0^n,\right]\in \mathcal{D}(\bm{x})\subset \mathbb{R}^d$ that captures the important $\bm{x}$-variations of $\bm{f}(\bm{x})$, and compute corresponding low-fidelity outputs $X^L_t = \left[\bm{f}^L(\bm{x}_0^1),\, \cdots\,,\bm{f}^L(\bm{x}_0^n),\right]$,
    \item out of the evaluated low-fidelity set $X^L_t$, select the set of interpolation nodes $\{\bm{f}^L(\bm{x}_0^i)\}$, columns of $X^L_t$, and compute~$\{\bm{f}^H(\bm{x}^i_0)\}$,
    \item interpolate the realizations $\{\bm{f}^H(\bm{x}_0^i)\}$ to $\bm{f}^L(\bm{x})$ by using the low-fidelity samples to dictate the topology on~$\mathbb{R}^m$.
\end{enumerate}
The low-fidelity model is used at three different moments: first, to inform the choice of the interpolation nodes; then, to shape the interpolating function, and third, to locally evaluate the interpolating function. The key advantage over other interpolation methods is that the approximate topology of the output can be exploited; this fact helps mitigate the curse of dimensionality. Subsections \ref{sub:interNodes} and \ref{sub:modelSynth} report the methodology developed by Narayan et al. \cite{narayan_2014}. Details that are not relevant to the problem of this work are neglected. Subsection~\ref{sub:mfmaneuvers} describes how the method has been adapted to maneuvering target trajectories, and is one of the main contributions of this paper.

\subsection{Interpolation Nodes Selection}
\label{sub:interNodes}
Let the low-fidelity distance between a vector $\bm{x}_0\in \mathbb{R}^m$ and a subspace $\mathbb{X}\subset \mathbb{R}^m$ be:
\begin{equation}
    d^L\left(\bm{x}_0,\mathbb{F}\right) = \inf_{\bm{w}\in \mathbb{X}_t}\|\bm{f}^L(\bm{x}_0)-\bm{f}^L(\bm{w})\|.
\end{equation}
The selection of the interpolating nodes is iterative and such that the next selected point is the one whose distance from the subspace spanned by the set of the previously selected points is maximum.
The nodes ${\bm{\gamma}}^1,...\, ,{\bm{\gamma}}^r$, where $r$ is the number of interpolation nodes, are a subset of the columns of $X_t^L$ and are greedily selected:
\begin{equation}
\label{eq:impSamples}
    \bm{\gamma}^i =\arg\max_{(\bm{x}_t^L)\in X_t^L} d^L\left(\bm{x}_t^L,\{\bm{\gamma}^1, \dots,\bm{\gamma}^{i-1}\}\right).
\end{equation}
Note that for tractability the domain $\mathbb{X}_t$ has now been discretized.
This minmax problem can be solved via a single iteration of the pivoted Cholesky decomposition~\cite{harbrecht2012low,bebendorf2003adaptive} of the Gram matrix $G^L=(X^L_t)^TX^L_t$:
\begin{equation}
(X^L_t)^TG^L X^L_t = PLL^TP.    
\end{equation}
The pivoted decomposition is operated one pivot at a time; each pivoting solves one iteration of the problem~\eqref{eq:impSamples}. The decomposition is partial because only the first $r$ permutations need be computed. The iterative pivoted Cholesky decomposition ensures that all elements of $\{\bm{\gamma}^1, \dots,\bm{\gamma}^{r}\}$ are linearly independent, which is a requirement for model synthesis. The computation of the first $r$ rows and columns of the Cholesky factor $L$ of the pivoted Gramian $G^L$, needed for model synthesis, is a byproduct of the decomposition. For a step-by-step algorithm, see, \textit{e.g.},~\cite{jones_2019a}.

\subsection{Model Synthesis}
\label{sub:modelSynth}
First, by exploiting the entries of $X^L_t$, an interpolation operator is constructed. Then, the operator is applied with the data $\bm{f}^H(\{\bm{\gamma}^1, \dots,\bm{\gamma}^{r}\}):=\Gamma^H_t$. Let $\bm{f}^L(\bm{x}_0)$ be a point-wise low-fidelity solution that does not need to be part of the previously included data set; the coefficients $c_n$ are determined by solving 
\begin{equation}
    \left\langle\sum_{n=1}^N c_n \bm{f}^L(\gamma^n),\bm{\phi}\right\rangle = \left\langle \bm{f}^L(\bm{x}_0),\bm{\phi}\right\rangle,\qquad\forall\bm{\phi}\in \bm{f}^L(\{\bm{\gamma}^1, \dots,\bm{\gamma}^{r}\}),
\end{equation}
which implies
\begin{equation}
\label{eq:coeffs}
    \mathbf{c} = \left(L L^T\right)^{-1}\mathbf{g},
\end{equation}
where $\mathbf{c}$ is the stacked vector of the coefficients $c_n$, and $\mathbf{g}$ is the stacked vector of $\left\langle \bm{f}^L(\bm{x}_0),\bm{\phi}\right\rangle$.
Then, the interpolation is applied with the high-fidelity data $\Gamma^H_t$:
\begin{equation}
\label{eq:interpolation}
    \hat{\bm{x}}^H = \sum_{n=1}^N c_n \bm{f}^H(\gamma^n).
\end{equation}
Since the pivoting matrix is known before performing the last computation, only the columns corresponding to the first $r$ pivots of $F^H$ need to be computed.
In practice, a linear interpolator is constructed using the low-fidelity topology and data, but then the evaluation is done using the high-fidelity data. The error introduced by the interpolation can then be quantified~\cite{zucchelli2021multi}, and it is used to estimate additional process noise in the filter.

\subsection{Multi-fidelity Stochastic Collocation for Maneuvering Spacecraft}
\label{sub:mfmaneuvers}
\begin{figure*}[htb]
    \centering
     \includegraphics[trim= 30mm 30mm 70mm 5mm, clip,width=3.25in]{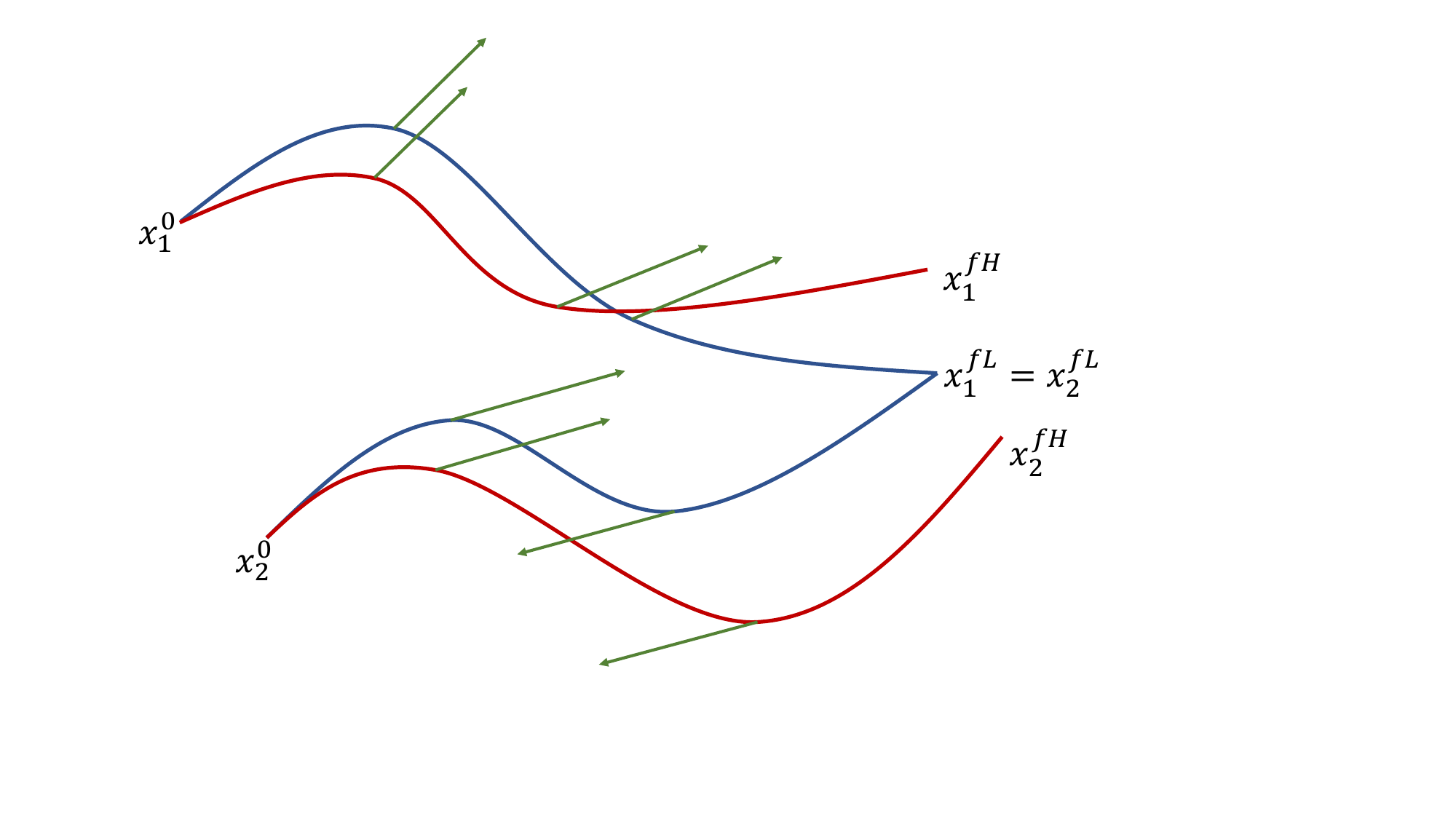}
    \caption{Example of problematic multi-fidelity maneuvering spacecraft trajectories.}
    \label{fig:MF_maneuvering}
\end{figure*}

In previous applications of the multi-fidelity stochastic collocation to astrodynamics problems, only the final state was considered by the interpolation. Thus, maneuvering trajectories ending at the same state may have the same correction, even if they started from different initial states and had different thrust profiles.
These trajectories would require different corrections from one another, since the corresponding propagations are different. Figure~\ref{fig:MF_maneuvering} illustrates an example of spacecraft trajectories ending at the same state when propagated with the low-fidelity model (blue) and at different states when propagated with the high-fidelity model (red). For simplicity, the thrust (green) is drawn as impulsive instead of continuous. 
To prevent this issue, the states at the start and end of every thrust segment are interpolated by the multi-fidelity method. The algorithm now effectively interpolates over $6+6\,n_{\Delta V}$ variables, with $n_{\Delta V}$ the number of thrust segments, causing an increase in the required number of high-fidelity samples. In this case, the increase in number of high-fidelity samples is sublinear in the number of variables.

\section{Thrust Profile Distribution}
\label{sec:thrustdistr}
\begin{figure*}[htb!]
    \centering
    \includegraphics[trim= 0mm 86mm 6mm 89mm, clip,width=3.25in]{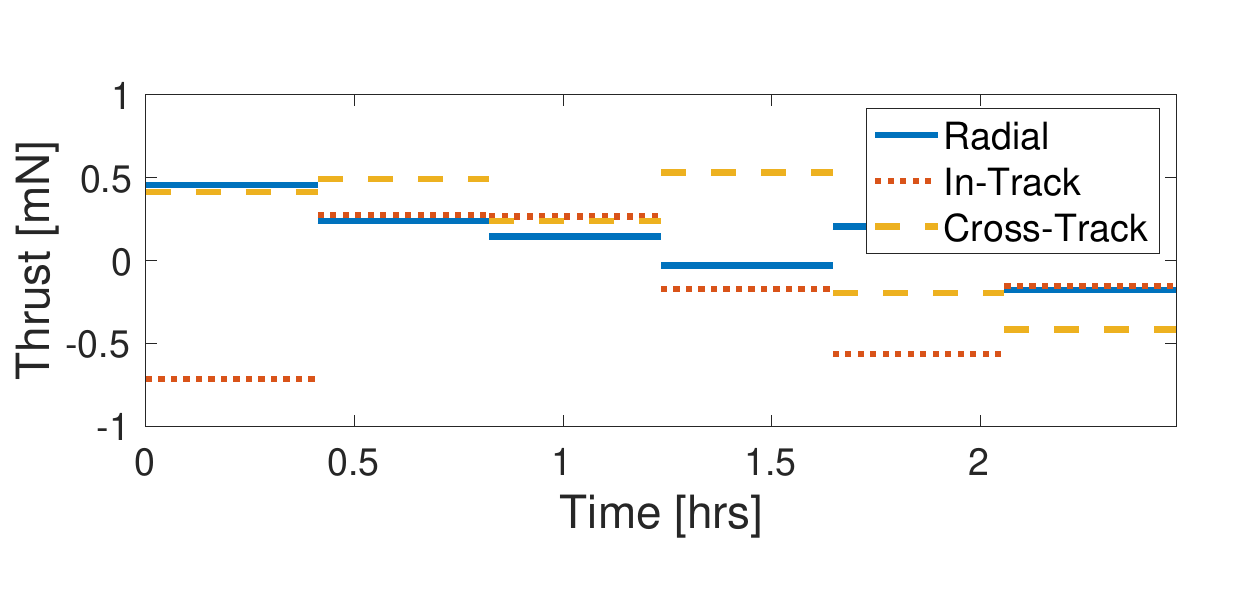}
    \caption{Example of thrust profile sample between two measurement passes.}
    \label{fig:RICThrust}
\end{figure*}

This section is one of the main contributions of the paper, as it describes and justifies the selection of the distribution for the thrust profiles. The thrust profile distribution is the model used for process noise in the filter.
A real thrust profile has infinite dimensions, since the thrust at every instant is a free variable; to keep the problem tractable, the profiles are assumed piece-wise constant in the radial, in-track, and cross-track~(RIC) frame. Segments of constant thrust are preferred over thrust Fourier coefficients~\cite{ko2014essential} because they provide a direct correlation between magnitude and $\Delta V$.
This fact enables a sampling method where there is a strict relation between probability density and $\Delta V$. An example of a thrust sample $\bm{t}$ made of six segments is depicted in Fig.~\ref{fig:RICThrust}.

The thrust profile is assumed to be distributed according to a multivariate Laplace~distribution, which is super-Gaussian.
The probability density function~(p.d.f.) of a multivariate Laplace~distribution with mean~$\bm{0}$ and covariance $Q$ evaluated at $\bm{t}$ is:
\begin{equation}
    f_{ML}(\bm{t};\bm{0},Q)=\frac{2}{\sqrt{\left|2\pi Q\right|}}\left(\frac{\bm{t}^TQ^{-1}\bm{t}}{2}\right)^{v/2}K_v\left(\sqrt{2\bm{t}^TQ^{-1}\bm{t}}\right),
\end{equation}
where $v=\left(2-d\right)/2$, the dimension of $\bm{t}$ is $d=3\,n_{\Delta V}$, and $K_v\left(\cdot\right)$ is the modified Bessel function of the second kind. Let the $A$-norm of a vector $\bm{t}$ be $\|\bm{t}\|_A \mathrel{\mathop:}=\sqrt{\bm{t}^TA\bm{t}}$. By properties of $K_v\left(\cdot\right)$, for $\|\bm{t}\|_{Q^{-1}} \gg 1$ (which occurs when the thrust is larger than several standard deviations), the multivariate Laplace~distribution asymptotically behaves as
\begin{equation}
    \lim_{\|\bm{t}\|_Q^{-1}\to\infty} f_{ML}(\bm{t};\bm{0},Q) e^{\sqrt{2}\|\bm{t}\|_{Q^{-1}}}\left(\|\bm{t}\|_{Q^{-1}}\right)^{1-v}= c ,
\end{equation}
where $c$ is a constant. The joint posterior distribution of initial state $\bm{x}_0$ and process noise $\bm{t}$, given a measurement $\bm{y}$ and assuming a linearized problem with multivariate Laplace process noise, initial state $\bm{x}_0\sim\mathcal{N}\left(\bar{\bm{x}},P_0\right)$, Gaussian measurement noise $\epsilon\sim\mathcal{N}\left({\bm{0}},R\right)$, and where the innovation distribution is assumed Gaussian, can be approximated by the following log~probability distribution when $\|\bm{t}\|_{Q^{-1}}$ is large:
\begin{align}
\label{eq:logprobability}
    \log p(\bm{x}_0,\bm{t}|\bm{y})\approx &-\frac{1}{2}\left(\bm{x}_0-\bm{\bar{x}}_0\right)^TP_0^{-1}\left(\bm{x}_0-\bm{\bar{x}}_0\right) - \sqrt{2}\|\bm{t}\|_{Q^{-1}} +\\
\nonumber    -&\frac{1}{2}\left(\bm{h}\left(\bm{x}_f\right)-\bm{y}\right)^T\left(HP_fH^T+R\right)^{-1}\left(\bm{h}\left(\bm{x}_f\right)-\bm{y}\right)+{(1-v)}\log \left(\|\bm{t}\|_{Q^{-1}}\right)+C,
\end{align}
where $C$ is a normalizing constant, $H$ is the measurement Jacobian, $P_f$ is the propagated state covariance, and $\bm{x}_f = \bm{f}(\bm{x}_0,\bm{t},t_f)$.
Let $J_t$ be the part of the probability distribution in Eq.~\eqref{eq:logprobability} associated with the thrust:
\begin{equation}
\label{eq:logthrust}
    \log J_t(\bm{t}) = - \sqrt{2}\|\bm{t}\|_{Q^{-1}}+{(1-v)}\log \left(\|\bm{t}\|_{Q^{-1}}\right).
\end{equation}
The derivative of the logarithm is very small when $\|\bm{t}\|_{Q^{-1}}\gg 1$, and the term can thus be neglected. Assume now that $Q$ is a multiple of the identity matrix. For a single thrust segment, the log p.d.f. is now affine in the total $\Delta V$. According to Tsiolkovsky's equation, the actual cost of maneuvers is exponential in the $\Delta V$, and this formulation of the thrust's p.d.f. reflects that.
For multiple thrust segments, one still has that $\log \left(J_{t}\left(a\bm{t}\right)\right)\approx\log \left(a J_{t}\left(\bm{t}\right)\right)$. The asymptotically exponential cost in $\Delta V$ is indeed the main reason why a multivariate Laplace distribution is preferred over the Gaussian distribution, or any other super-Gaussian distributions. 
The process noise variance $Q$ is still a parameter to be tuned, but thanks to the quasi linearly exponential dependency in the $\Delta V$, the filter is much less sensitive to the selection of $Q$ than a Gaussian filter would be.

In conclusion, the use of a multivariate Laplace distribution for the thrust profiles of the spacecraft allows to have a robust filter, mostly insensitive to the set standard deviation, and a p.d.f. that resembles more closely the actual cost of operating a spacecraft, all while including time-varying thrust within the same measurement gap.

\section{Rare Event Simulation Particle Filter}
\label{sec:rareevent}
This section describes the rare event simulation particle filter, one of the main contributions of this paper. When unexpectedly large deviations from the assumed model occur, the proposed approach offers greatly reduced bias and variance with respect to a standard particle filter, thanks to a better representation of the tails of the distributions. As a maneuver can often cause large deviations, such a feature is a key component of the overall algorithm described in this paper. This section provides theoretical and empirical results.

\subsection{Rare Event Simulation for State Estimation}
This subsection shows how concepts from rare event simulation, and specifically importance sampling, can be adapted to reduce the variance and the bias of a particle filter estimate when a large deviation, or maneuver, occurs.

In statistical estimation one generally wishes to obtain the minimum mean square error~(MMSE), equal to the following normalized integral:
\begin{equation}
\label{eq:mmse}
   {x}^{MMSE} = \mathbb{E}_{p(x|y)}[x] = \frac{\mathbb{E}_{p(x)}[xp(y|x)]}{\mathbb{E}_{p(x)}\left[p(y|x)\right]}=\frac{\int_{\mathcal{D}(x)} x p(y|x)p(x)dx}{\int_{\mathcal{D}(x)} p(y|x)p(x)dx}:=\frac{A(y)}{B(y)},
\end{equation}
where $p(x)$ is the prior distribution of $x$, $p(y|x)$ is the measurement distribution conditioned on the state, and $p(x|y)$ is the posterior distribution. The last equality of Eq.~\eqref{eq:mmse} is a definition. Note that both $A$ and $B$ are a function of $y$, a random variable; however, given the data, they are deterministic. For brevity, the dependency on $y$ is dropped. The following analysis is carried out for scalar state and measurement, but can be easily generalized to arbitrary dimensions.
A particle filter is equivalent to using crude Monte Carlo to approximate both numerator and denominator of Eq.~\eqref{eq:mmse}:
\begin{equation}
    \hat{x}^{MMSE}=\frac{\sum_{i=1}^N \frac{1}{N}x_i p(y|x_i)}{\sum_{i=1}^N \frac{1}{N} p(y|x_i)}=\frac{\hat{A}}{\hat{B}},
\end{equation}
where $x_i\sim p(x)$, and $\hat{\cdot}$ symbolizes an empirical estimate.
The expected value and the covariance of a ratio of two random variables~\cite{elandt1980survival} can be approximated by Taylor series:
\begin{equation}
\label{eq:exvratio}
    \mathbb{E}_{p(x)}\left[\frac{\hat{A}}{\hat{B}}\right] \approx
    \frac{A}{B}\left(1-\frac{\text{Cov}_{p(x)}\left[{\hat{A},\hat{B}}\right]}{A B} + \frac{ \text{Var}_{p(x)} \left[{\hat{B}}\right]}{B^2}\right),
\end{equation}
\begin{equation}
\label{eq:ratiovariance}
    \text{Var}_{p(x)}\left[\frac{\hat{A}}{\hat{B}}\right]\approx\left(\frac{A}{B}\right)^2\left(\frac{\text{Var}_{p(x)}\left[{\hat{A}}\right]}{A^2}-2\frac{\text{Cov}_{p(x)}\left[{\hat{A},\hat{B}}\right]}{A\,B}+\frac{\text{Var}_{p(x)}\left[{\hat{B}}\right]}{B^2}\right).
\end{equation}
From Eqs.~\eqref{eq:exvratio}-\eqref{eq:ratiovariance} one can compute the expected \ac{rmse} of the estimate:
\begin{equation}
\label{eq:RMSE}
    \mathbb{E}_{p(x)}\left[RMSE(\hat{x})\right] = \sqrt{\left(\frac{A}{B}-\mathbb{E}_{p(x)}\left[\frac{\hat{A}}{\hat{B}}\right]\right)^2 + \text{Var}_{p(x)}\left[\frac{\hat{A}}{\hat{B}}\right]},
\end{equation}
Note that this \ac{rmse} is the expectation over the sampling distribution, and refers to the error between the computed MMSE and the exact MMSE; the actual error of the estimate might in fact be larger.

The bootstrap particle filter is also called sequential importance sampling~(SIS) because it exploits a sampling distribution, the state's prior $p(x)$, to compute an expected value for a different distribution, $p(x|y)$. Importance sampling for particle filters is necessary because $p(x|y)$ is unknown, and $p(x)$ is just considered a good guess. However, importance sampling in the framework of rare event simulation is different, and consists of a variance reduction technique~\cite{rubino2009rare,beck2015rare}. In this context, some instances of importance sampling are also called exponential twisting, or exponential tilting.

Similar to what is normally done to compute the probability of a rare event, it is here proposed to sample from a proposal distribution $q(x)$, strictly positive for every $x$ where the posterior is non-zero.
The samples are then given the weights
\begin{equation}
\label{eq:importanceSampling}
    w_i(x_i) = \frac{p(x_i)}{q(x_i)}.
\end{equation}
Because of properties of importance sampling, all the first raw moments of $\hat{A}$ and $\hat{B}$ computed empirically are unaffected by the choice of~$q(x)$, but all second moments are. While rare event simulation generally only reduces the variance of the estimate, in this case it has an effect on the bias too, because of Eq.~\eqref{eq:exvratio}.
The second moments of $\hat{A}$ and $\hat{B}$ can be computed:
\begin{align}
    \text{Var}_{q(x)}\left[\hat{A}\right]   &=\frac{1}{N} \int_{\mathcal{D}(x)} \left(x p(y|x)-A\right)^2 \frac{p^2(x)}{q(x)} dx,\\
    \text{Cov}_{q(x)}\left[\hat{A},\hat{B}\right]   &= \frac{1}{N} \int_{\mathcal{D}(x)} \left(x p(y|x)-A\right)\left(p(y|x)- B\right) \frac{p^2(x)}{q(x)} dx,\\
    \text{Var}_{q(x)}\left[\hat{B}\right] &= \frac{1}{N} \int_{\mathcal{D}(x)} \left(p(y|x)- B\right)^2 \frac{p^2(x)}{q(x)} dx.
\end{align}

From these equations and Eqs.~\eqref{eq:exvratio}-\eqref{eq:RMSE}, opportunely computing expectations over $q(x)$ instead of $p(x)$, one may select $q(x)$ to reduce the expected \ac{rmse} of the estimate.
It is a known result that the variance of $\hat{B}$ is minimized by sampling from the posterior distribution. Sampling from the posterior distribution is what many inference techniques, such as the auxiliary particle filter~\cite{pitt1999filtering} and Markov Chain Monte Carlo methods~\cite{metropolis1953equation}, attempt to do. Because they require having an associated measurement before propagation, neither of the above mentioned techniques are utilized in this work.

\subsection{Rare Event Simulation Filter with Laplace Prior}
Suppose now that one wishes to estimate $x\in \mathbb{R}$, which has a Laplace prior distribution $p(x)$ with zero mean and rate parameter $\lambda_L$:
\begin{equation}
    p(x) = \frac{\lambda_L}{2}e^{-\lambda_L|x|}.
\end{equation}
The corresponding standard deviation is $\sigma_L = \sqrt{2}/\lambda_L$.
The measurement likelihood function $\Lambda(y,x)$ is Gaussian with standard deviation $\sigma_y$ and mean~$x\gg \lambda_L\sigma_y^2$.
The numerator of Eq.~\eqref{eq:mmse} can be approximated by removing the absolute value from the Laplace distribution:
\begin{equation}
    B=\mathbb{E}_{p(x)}\left[\Lambda(y,x)\right] = \int_{-\infty}^\infty \frac{1}{\sigma_y\sqrt{2 \pi}} e^{-\frac{(x-y)^2}{2\sigma_y^2}} \frac{\lambda_L}{2} e^{-\lambda_L|x|}dx \approx \frac{\lambda_L  e^{\frac{1}{2}\lambda_L\left(\lambda_L\sigma_y^2-2y\right)}}{2},
\end{equation}
The assumption is valid for large $y$, since the Gaussian tail decays much faster than the Laplace tail. The same assumption is used in the computation of all the integrals of this section. The MMSE can then be approximated too:
\begin{equation}
    \mathbb{E}_{p(x|y)}\left[x\right]\approx y-\lambda_L\sigma_y^2.
\end{equation}

By drawing $N$ samples from $q(x)=\frac{\Tilde{\lambda}_L}{2}e^{-\Tilde{\lambda}_L|x|}$, where $\Tilde{\lambda}_L$ is a variable to be tuned, the variance of the estimate $\hat{B}$ is:
\begin{align}
\label{eq:denVar}
\text{Var}_{q(x)}\left[\hat{B}\right]=\frac{1}{N}\text{Var}_{q(x)}\left[\Lambda(y,x)\frac{p(x)}{q(x)}\right]&=\frac{1}{N}\int_{-\infty}^\infty \Lambda^2(y,x)\frac{1}{2}\lambda_L\frac{e^{-\lambda_L|x|}}{\Tilde{\lambda}_L^2e^{-2\Tilde{\lambda}_L|x|}}dx
    -B^2
    \approx\\
    \nonumber &\approx \frac{1}{N}\frac{\lambda_L^2}{\Tilde{\lambda}_L} \frac{e^{\frac{1}{4}(\Tilde{\lambda}_L-2\lambda_L)(\sigma_y^2(\Tilde{\lambda}_L-2\lambda_L)+4y)}}{4\sqrt{\pi}\sigma_y}
    - B^2. 
\end{align}
The variance of $\hat{A}$ and the covariance of $\hat{A}$ and $\hat{B}$ can be approximated in a similar manner, and the computation is not reported for brevity.
Finally, the total expected \ac{rmse} sampling from $q(x)$ can be computed with Eqs.~\eqref{eq:exvratio}-\eqref{eq:RMSE}, but computing expectations over $q(x)$ instead of $p(x)$:
\begin{equation}
\label{eq:RMSEq}
    \mathbb{E}_{q}\left[RMSE(\hat{x})\right] = \sqrt{\left(\frac{A}{B}-\mathbb{E}_q\left[\frac{\hat{A}}{\hat{B}}\right]\right)^2 + \text{Var}_q\left[\frac{\hat{A}}{\hat{B}}\right]},
\end{equation}
which is therefore a function of $\lambda_L,\Tilde{\lambda}_L,y$ and $\sigma_y$. After observing $y$ and by opportunely selecting $\Tilde{\lambda}_L$, one can therefore adjust the second moments of the joint distribution of $\hat{A}$ and $\hat{B}$ such that the \ac{rmse} is minimized.

\subsection{Results}
\label{sub:resPFresults}
\begin{figure*}[htb!]
    \centering
    \includegraphics[trim= 12.5mm 65mm 14mm 70mm, clip,width=3.2in]{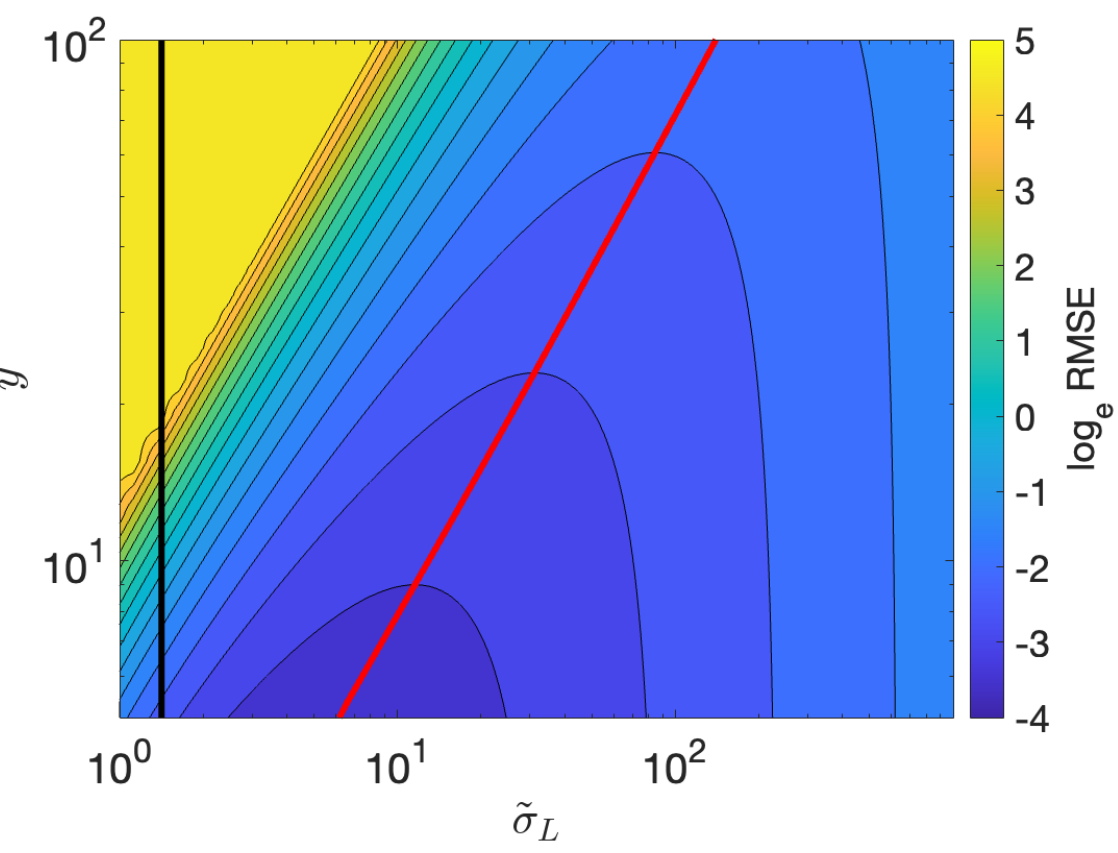}
    \caption{Expected logarithm of the RMSE according to Eq.~\eqref{eq:RMSEq}.}
    \label{fig:theoreticalPF}
\end{figure*}
Figure~\ref{fig:theoreticalPF} shows the theoretical expected log \ac{rmse} of the estimate, computed as in Eq.~\eqref{eq:RMSEq}, obtained with a particle filter with 2,500 samples when $\lambda_L=1$ and $\sigma_y=1$. The \ac{rmse} is plotted as a function of $\Tilde{\sigma}_L=\sqrt{2}/\Tilde{\lambda}_L$ and of the measurement $y$. The plot is capped when log \ac{rmse} is larger than 5 because the approximation can get inaccurate. When $q(x)=p(x)$, marked as the vertical black line, the estimate quickly degrades for growing values of $y$. However, using the optimal $\Tilde{\sigma}_L$, the red line in the figure, the decrease in accuracy is mitigated. Note that the optimal value of $\Tilde{\sigma}_L$ is a function of $y$; therefore, if one has to sample before knowing the value of $y$, it is best to sample from a mixture of distributions with multiple values of $\Tilde{\lambda}_L$. The optimal value of $\Tilde{\sigma}_L$ is very close to linear with $y$, with an empirical regression factor close to $\Tilde{\sigma}_{L,opt} = \sqrt{2}y$.

\begin{figure*}[htb!]
    \centering
    \subcaptionbox{1-dimensional prior}{\includegraphics[trim= 12mm 65mm 14mm 70mm, clip,width=3.2in]{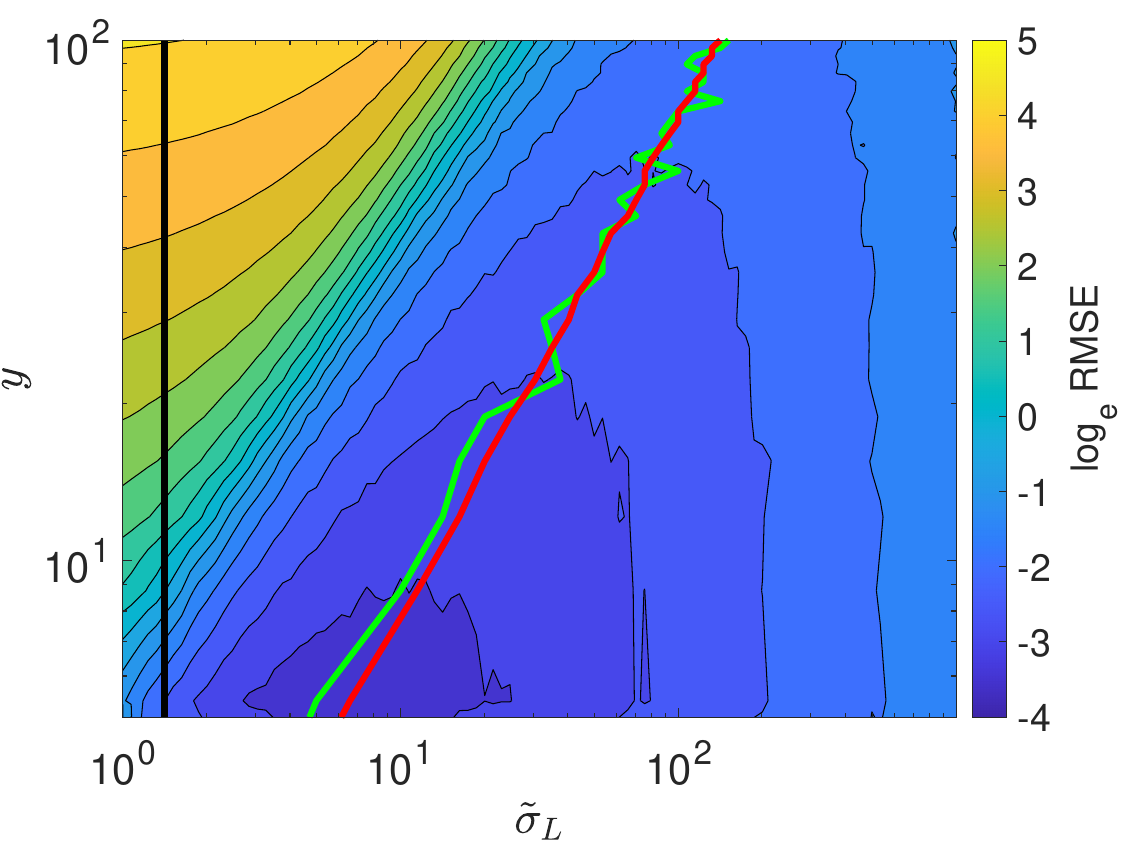}
    }
    \subcaptionbox{6-dimensional prior}{\includegraphics[trim= 12mm 65mm 14mm 70mm, clip,width=3.2in]{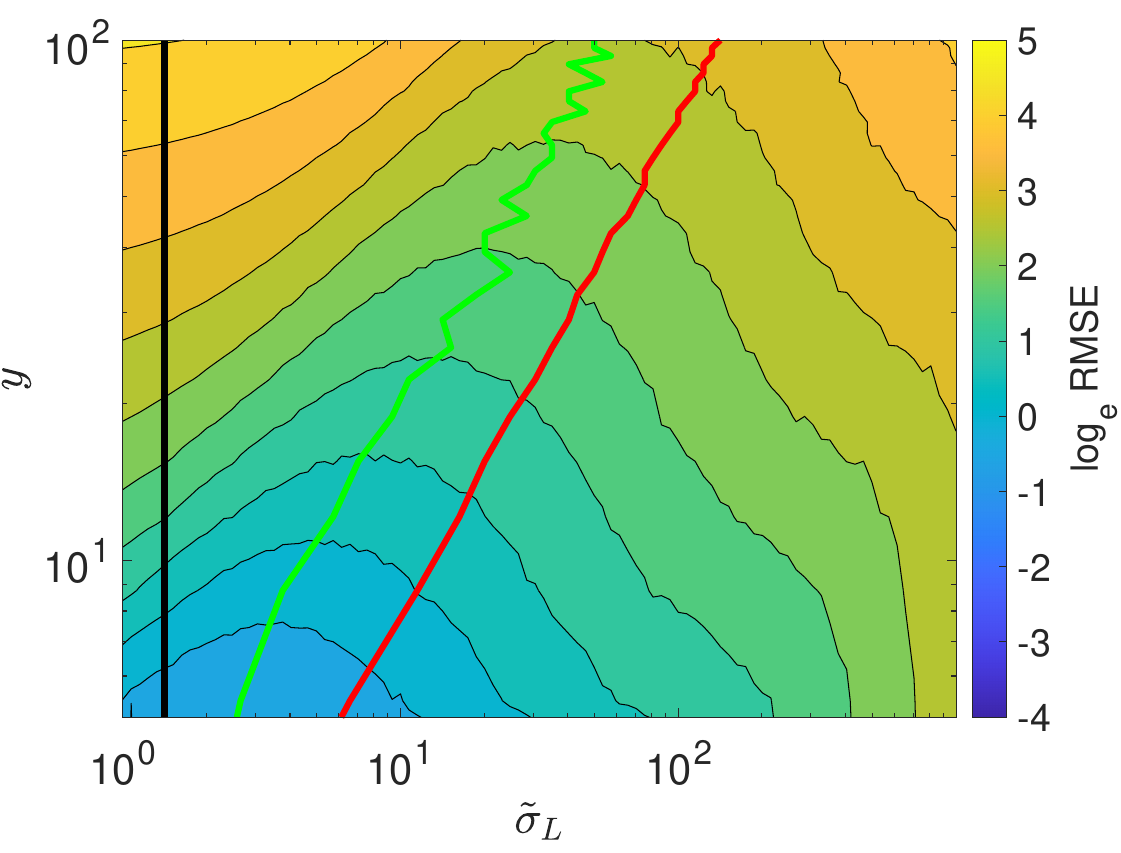}}
    \caption{RMSE for Monte Carlos of particle filters.}
    \label{fig:rmsemcpf}
\end{figure*}

Figure~\ref{fig:rmsemcpf} shows the same quantity of interest as Fig.~\ref{fig:theoreticalPF} with same color scale and contour lines, but computed empirically. Every data point provides the \ac{rmse} obtained from 1,000~particle filters, each with 2,500 particles. The left plot is for the one-dimensional case, and the right plot is for a 6-dimensional prior, with one dimensional measurement. For the latter, only the observed dimension is considered in the \ac{rmse}.
The red line is the same that minimizes the \ac{rmse} of Eq.~\eqref{eq:RMSEq}, and the green line minimizes the particle filters' \ac{rmse}.
Despite the noise, one can qualitatively verify that the red and green lines have similar slopes, and, in the one-dimensional case, they tend to overlap. Moreover, the similarity between the left plot in Fig.~\ref{fig:rmsemcpf} and Fig.~\ref{fig:theoreticalPF} extends to the colors and contour lines, implying that, as expected, the values are similar between the two figures. The similarity decreases moving toward the top-left corner of the plots.
Based on the right plot in Fig.~\ref{fig:rmsemcpf},with higher number of dimensions the impact of importance sampling is reduced and the optimal $\Tilde{\sigma}$ is smaller, but the benefits of using rare event simulation remain.

In conclusion, this section shows that a) with Laplace priors, it is necessary to use wider sampling distributions if large deviations occur, as is likely in maneuvering target tracking problems, and b) the optimal choice of distribution is a function of the measurement. This fact introduces a problem that is solved in Sec.~\ref{sub:pnIS}.

\section{Nearest-Neighbor Ensemble Gaussian Mixture Filter}
\label{sec:nnengmf}
This section introduces an ensemble Gaussian mixture filter~(EnGMF) aided by a $k$-nearest neighbor algorithm. An EnGMF~\cite{liu2016efficient} has several benefits over particle filters, most importantly the fact that it allows for reduced variance in the weights of the updated components, and thus reduces the risks of particle depletion and degeneracy. In an EnGMF, one samples particles from an initial distribution; the particles are propagated, and regularized into Gaussian kernels, forming together a \ac{gmm}. The measurement update then follows that of a Gaussian sum filter. At the following time-step, new particles are sampled from the updated \ac{gmm}. In the context of this work, the development of an EnGMF with kernels dependent on their neighboring samples is necessary because of the large diversity in the distance between sampled particles; therefore, using the same kernel for all particles may lead to small isolated components. The EnGMF is equivalent to the pre-regularized particle filter~\cite{hurzeler1998monte,oudjane2000progressive}, except that the measurement update is performed with a bank of weighed Kalman filters. Regularization consists of substituting Dirac delta functions with kernels, making the distribution continuous rather than discrete. In an EnGMF, the choice of kernel is limited to Gaussian distributions, with mean equal to the state of the corresponding particle. The only design variable left is the covariance matrix: 
\begin{equation}
    \nonumber\sum_{j=1}^N w_j\delta(\mathbf{x}_j,\mathbf{x}) \rightarrow \sum_{j=1}^N w_j\mathcal{N}\left(\mathbf{x};\mathbf{x}_j,\Sigma_j\right).
\end{equation}
Regularization modifies the original distribution and inflates the overall covariance. This inflation is considered an acceptable drawback that is outweighed by the fact that a discrete distribution is now continuous, and thus closer to the actual distribution under many other metrics, such as the Kullback-Leibler divergence.
A commonly used regularization method for the EnGMF is Silverman's rule of thumb~\cite{silverman1986density,liu2016efficient}: the kernel is Gaussian and the variance is a scaled empirical covariance.
Silverman's regularization is optimal for Gaussian distributions, and can be used for non-Gaussian distributions too. While it has been used to solve astrodynamics problems before~\cite{yun2022kernel}, when applied to this problem it fails, because with multivariate Laplace process noise the average distance between particles varies widely when moving from the mode to the tails. Using the same variance for all particles causes the density to either have undesired gaps in the distribution, or to have kernels so diffuse that too much information is lost.
For this reason, a locally adaptive and anisotropic kernel is chosen.

Starting from the approach proposed by Brox et al.~\cite{brox2007nonparametric}:
\begin{align}
\label{eq:brox}
    P_{k+1|k}^i &= \eta \mathbb{I} +\sum_{j=1}^{N}K_0\left(\bm{x}^i,\bm{x}^j\right)\frac{\left(\bm{x}^i-\bm{x}^j\right)\left(\bm{x}^i-\bm{x}^j\right)^T}{k-1},\\
    K_i&=\mathcal{N}\left(\bm{x};\bm{x}_i,P_{k+1|k}^i\right)
\end{align}
where $K_0$ is a particle-independent isotropic weighing function, $\eta$ is a regularization parameter, and $K_i$ is the kernel applied to particle $\bm{x}_i$. This method alone allows for anisotropic kernels that are dependent on the local orientation of the density, as opposed to anisotropic, particle independent kernels such as the Silverman's rule of thumb, or anisotropic, particle-dependendent kernels such as the classic version of the $k$-nearest neighbor kernel density estimator, \textit{e.g.},~\cite{scott2015multivariate}. The method from Eq.~\eqref{eq:brox} still uses a particle-independent, Gaussian function $K\left(\cdot\right)$ to weigh the particles contributions to the covariance of the kernel. This fact may lead to very small weights in the tails, where the particles are very far from one another, causing degeneracy of the kernel. Furthermore, the computational cost of computing the covariance for all kernels grows with $n^2$, since every particle contributes to the covariance of every other kernel. Hence, an improved version of the method is proposed.

\subsection{Proposed Approach}
Building on Eq.~\eqref{eq:brox}, substitute the particle independent kernel $K_0(\cdot)$ with a particle-dependent gating function $G_i(\bm{x}^i,\cdot)$. This substitution solves both of the above mentioned problems. The gating function is chosen to be based on $k$-nearest neighbors, such that
\begin{align}
\label{eq:gating}    G_i\left(\bm{x}^i,\bm{x}\right) &= \systeme{
  1  \qquad \text{if}\;    \|\bm{x}^i-\bm{x}\|    \leq h^i,
  0  \qquad \text{if}\;    \|\bm{x}^i-\bm{x}\|    > h^i},\\
  \label{eq:ensemblecovariance}
    P_{k+1|k}^i & = \sum_{j=1}^{N} G_i\left(\bm{x}^i,\bm{x}^j\right) \frac{\left(\bm{x}^i-\bm{x}^j\right)\left(\bm{x}^i-\bm{x}^j\right)^T}{k-1},\\
    K_i&=\mathcal{N}\left(\bm{x}; \bm{x}_i,P_{k+1|k}^i\right)
\end{align}
where $h^i$ is the distance between $\bm{x}^i$ and its $k$\textsuperscript{th} nearest-neighbor.
For problems with small dimensionality, $k$-dimensional trees can be built with cost $n\log n$, and each query costs $\log n$, making the overall regularization cost~$\propto c_1 kn+c_2 n \log n$, more manageable than the original approach of Eq.~\eqref{eq:brox}. As opposed to more traditional $k$-nearest neighbor kernel density estimators, here the value $h^i$ is not used as a scale distance, but it is used as a gating function. This way, instead of spherical kernels, kernels that better follow the data are allowed. The corresponding difference in outcomes will be shown in Sec.~\ref{sub:tpres}. The proposed regularization is equivalent to a sample covariance, centered on the given particle, and where samples are truncated after the $k$\textsuperscript{th} farthest one.
This anisotropic adaptive method has two main benefits over most regularization methods: first, the kernels are smaller where the data are denser, and larger where the data are less dense. This fact allows the estimator to more closely approximate the data when it is dense, while avoiding undesired gaps. Second, instead of being spheres, or ellipses oriented along predefined axes, the kernels follow the local orientation of the data. A paper describing an alternative approach for an adaptive EnGMF was presented by Popov and Zanetti~\cite{popov2023ensemble} shortly before this work was submitted for peer-review (the method described in this section was developed independently from their paper). In that work, the function $G_i(\bm{x}_i,\bm{x})$ is no longer a gating function, but it is a Gaussian kernel with covariance $h^i\,\mathbb{I}$. Consequently, the computational cost increases with $n^2$.

In the spacecraft tracking problem, the state includes position and velocity measures, which cannot be compared with one another.
The neighbors are therefore sought in dimensionless canonical units, which are
\begin{align}
    \bm{r}_{LU} & = \frac{\bm{r}}{R_e},\\
    \bm{v}_{LU} & = \bm{v}\sqrt{\frac{R_e}{\mu}},
\end{align}
where $R_e$ and $\mu$ are the planet's equatorial radius and gravitational parameter, respectively. If additional parameters were to be estimated, such as the drag coefficient, one should find suitable normalization constants, most likely based on how nonlinear are the effects of such parameters. 

The total prior p.d.f. can be computed as:
\begin{equation}
    p(\bm{x})=\sum_{i=1}^N \frac{w_i}{\sqrt{\left|2\pi P^i_{k+1|k}\right|}}\,\text{exp}\left({-\frac{1}{2}\left(\bm{x}-\bm{x_i}\right)^T\left(P^i_{k+1|k}\right)^{-1}\left(\bm{x}-\bm{x_i}\right)}\right).
\end{equation}
This approach allows for a $k$-nearest neighbor method that is anisotropic and that accounts for local correlation between variables, which are characteristics that are often missing in that category of methods. 

The combination between prior importance sampling and density-dependent regularization offers the additional advantage that covariance inflation is mitigated: the samples with the highest weights lie in the densest regions, where the $k$\textsuperscript{th} neighbor is closer than it would be in the less dense regions.

\section{Filter Architecture}
\label{sec:filtersummary}
This section synthesizes the contributions of the previous sections into a single filter.
As with most dynamic parameter estimation filters, this algorithm is divided into two main components: the construction of the transitional prior, and the measurement update. In this case, the transitional prior is constructed by sampling particles from a proposal density, weighing them according to importance sampling, propagating the particles, and regularizing. The update is made of two phases: first, the relative weights of all components are updated; second, an iterated Batch Least Squares~(iBLS) is used to update mean and covariance for each component. 
\subsection{Construction of the Transitional Prior}
\label{sec:transprior}
This subsection describes how the explicit transitional prior is approximated. Specific attention is given to reproducing the tails of the distribution.
By employing a Gaussian sum, as described in Sec.~\ref{sec:nnengmf}, the distribution's tails are bound to be sub-Gaussian. The objective of this section is to show how importance sampling described in Sec.~\ref{sec:rareevent} works in tandem with the nearest-neighbor ensemble Gaussian mixture filter presented in Sec.~\ref{sec:nnengmf}; together they provide a distribution that resembles a super-Gaussian distribution for a large portion of the domain. This result is achieved while representing the shape of the distribution as closely as possible, and inflating the total covariance as little as possible. The construction of the transitional prior consists of the following steps. For each particle:
\begin{enumerate}
    \item sample initial state from a Gaussian sum prior;
    \item sample a thrust profile from a proposal density distribution;
    \item weigh the particle according to its thrust profile;
    \item propagate the particle with its initial state and thrust profile; and
    \item perform regularization.
\end{enumerate}

\subsubsection{Process Noise Importance Sampling}
\label{sub:pnIS}

As explained in Sec.~\ref{sec:rareevent}, the proposed filter samples the process noise from a wider distribution than the one assumed. This way, deviations that are potentially orders of magnitude larger than the assumed standard deviation will still be represented.

Sampling from a multivariate Laplace distribution with mean $\bm{0}$ and variance $Q$ is done in two steps~\cite{kotz2001laplace}. First, a random sample is drawn from a normal distribution with mean $\bm{0}$ and variance $Q$. Then, all components of the drawn sample are multiplied by the square root of a sample drawn from an exponential distribution with scale 1:
\begin{equation}
\label{eq:MLsampling}
\bm{t} = \sqrt{\tau}\bm{r} \qquad \tau\sim Exp(1),
\end{equation}
where $\bm{r}\sim \mathcal{N}(0,{Q})$.
As in Sec.~\ref{sec:rareevent}, let $q\left(\bm{t}\right)$ be the sampling, or proposal, distribution, and $p\left(\bm{t}\right)$ be the target distribution. The weight $w_i$ can be computed using Eq.~\eqref{eq:importanceSampling}.
Although there are asymptotic guarantees of convergence whenever the domain of $q(\cdot)$ fully encompasses the domain of $p\left(\cdot\right)$, choosing too wide or too small a distribution can lead to very slow convergence when a finite number of samples are used (as was demonstrated in Sec.~\ref{sec:rareevent}).
To be conservative, a mixture of densities is chosen as the proposal distribution. Let $u\sim\mathcal{U}(0,1)$:
\begin{equation}
\label{eq:samplingMixture}
\bm{t} = \begin{cases}
    \sqrt{\tau}\bm{r}   \qquad\qquad \text{if}\, u\leq\alpha_0,\\
   a_i\sqrt{\tau}\bm{r} \,\,\, \,\,\,\,\qquad   \text{if}\, \sum_{l=0}^{i-1}\alpha_l< u \leq  \sum_{l=0}^{i}\alpha_l,\, i\geq 1,\\
\end{cases}
\end{equation}
where $\alpha_i$ are positive weights such that $\sum_{i=0}^{n}\alpha_i=1$, and $a_i>1$ for every $i$. In this case, the weight is computed as:
\begin{equation}
\label{eq:priorWeights}
    w_i = \frac{f_{ML}(\bm{t};\bm{0},\hat{Q})}{\alpha_0 f_{ML}(\bm{t};\bm{0},\hat{Q})+\sum_{l=1}^n(1-\alpha-\sum_{j=1,j\neq l}^n\alpha_j)f_{ML}(\bm{t};\bm{0},a_l^2\hat{Q})}.
\end{equation}
Sampling from a mixture of several multivariate Laplace distributions allows the filter to better react to a wider range of thrust magnitudes.
Using importance sampling for the thrust still requires the guess of an upper bound for the thrust magnitude. However the practitioner should not worry about exceeding it, because it only mildly affects the final estimate. The only drawback caused by using too large a distribution results from wasting some computational power on particles that have very low likelihood.

\subsubsection{Results}
\label{sub:tpres}
This subsection presents some illustrative results on the generation of the transitional. First, different importance sampling distributions are tested. Then, different regularization techniques are tested, using the same proposal density as in Eq.~\eqref{eq:samplingMixture}. The comparison is both visual and numerical: distributions whose tails decay slower and that follow the target distribution more closely are preferred. At the same time, for similar distributions, the one with the smaller uncertainty inflation is preferred.

\begin{figure*}[htb!]
    \centering
    \includegraphics[trim= 0mm 0mm 29mm 0mm, clip,height=.4\linewidth]{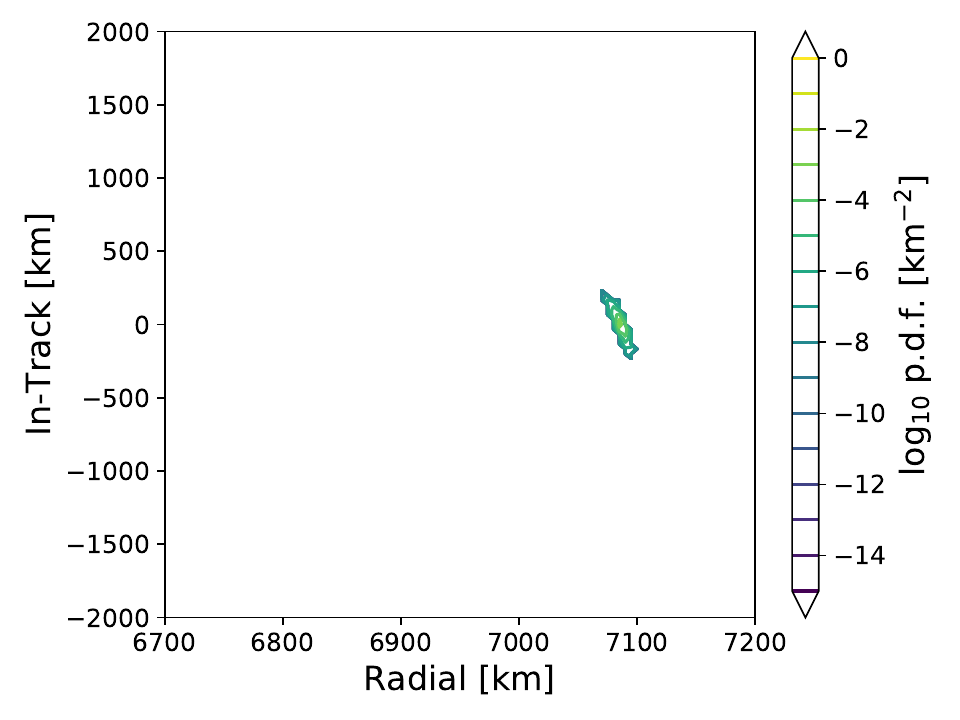}   \includegraphics[trim= 0mm 0mm 29mm 0mm, clip,height=.4\linewidth]{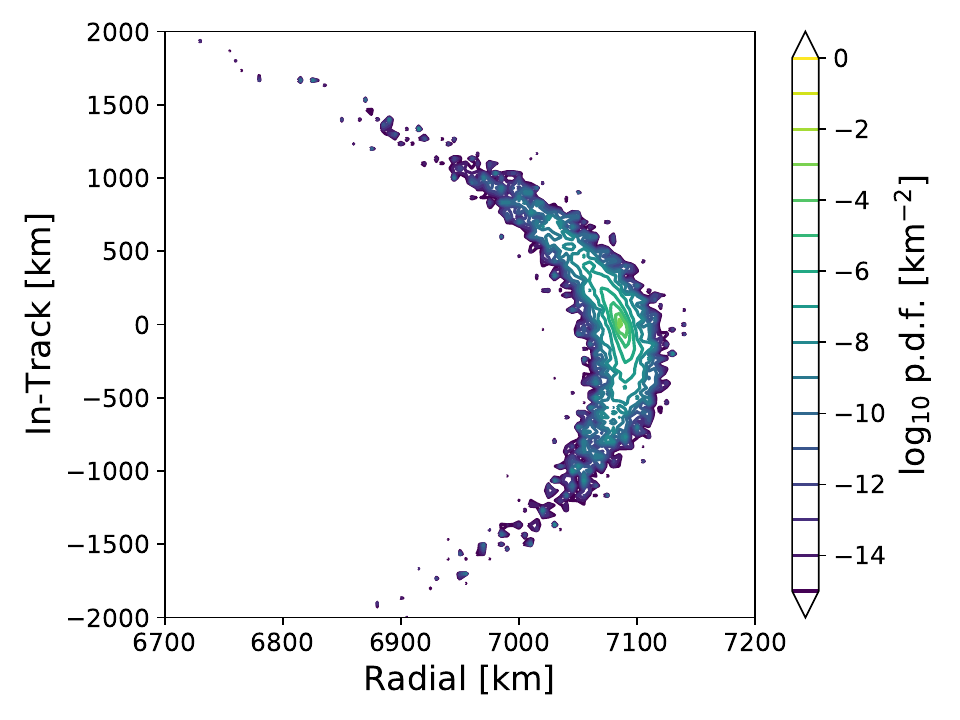} 
    \includegraphics[trim= 133mm 0mm 0mm 0mm, clip,height=.4\linewidth]{figparts_NT250000_NN1000_nx121.pdf} 
    \caption{Transitional prior without~(left) and with~(right) importance sampling.}
    \label{fig:transprior}
\end{figure*}

To accentuate the effect of the thrust, in every direction the initial uncertainty is 10~m in position and 10~mm/s in velocity. The initial state is
\begin{equation}
    \nonumber \left[-1,344~\text{km}\quad 4,909~\text{km}\quad 4,912~\text{km}\quad -7.375~\text{km/s}\quad -1.006~\text{km/s}\quad -1.010~\text{km/s}\right].
\end{equation}
The standard deviation on the acceleration, assumed constant during the propagation, is 50~\textmu m/s\textsuperscript{2} in each direction. The maximum standard deviation of the acceleration used in the proposal distribution is 1,600~\textmu m/s\textsuperscript{2}.
The propagation lasts 27,120~s, or approximately 7.53~hrs.
Fig.~\ref{fig:transprior} shows the explicit transitional prior obtained with and without importance sampling. The importance sampling is here performed with a mixture of distributions as in Eq.~\eqref{eq:samplingMixture}. Nine distributions are used, logarithmically spaced between the minimum and the maximum. The plots, which can be considered as a reference for the true distributions, are obtained by summing the weights of 250,000 propagated particles into the corresponding bins. The contour lines are in logarithmic scale, and end where the distribution's p.d.f. is less than 10\textsuperscript{-15} km\textsuperscript{-2}. It is clear from the figures that sampling from the original distribution fails to preserve the super-Gaussian tails. Table~\ref{table:stdevIS} reports the overall standard deviation obtained by the different importance sampling methods, using 2,500 particles and regularizing with 99 neighbors according to the method proposed in Sec.~\ref{sec:nnengmf}. The table shows that the proposed regularization moderately increases the overall uncertainty, and so does the importance sampling. Using only a large distribution for importance sampling causes excessive covariance inflation: the variance in the sample weights is much larger in that case, and a single sample receives 25\% of the total probability mass. This result is an additional motivation to use a mixture of distributions.

\begin{table}
\centering
\caption{Standard deviations using different proposal distributions, after time update.}\label{table:stdevIS}
	\begin{tabular}{cccccc}
		\hline
		 Importance Sampling  & Regularization & Pos. 1$\sigma$ [km] & Vel. 1$\sigma$ [m/s] \\ \hline
 \multirow{2}{*}{No}  & Yes & 2.820 & 3.117 \\
   & No & 2.0135 & 2.218 \\
  \multirow{2}{*}{Only largest distribution}  & Yes & 23.137 & 27.000 \\
   & No & 8.315 & 8.709\\
  \multirow{2}{*}{Mixture of distributions} & Yes & 5.315 & 6.069 \\
   & No & 2.893 & 3.086\\
   \hline
	\end{tabular}
\end{table}

\begin{figure*}[htb!]
    \centering
    \subcaptionbox{Single Gaussian}{\includegraphics[trim= 0mm 0mm 29mm 0mm, clip,height=.4\linewidth]{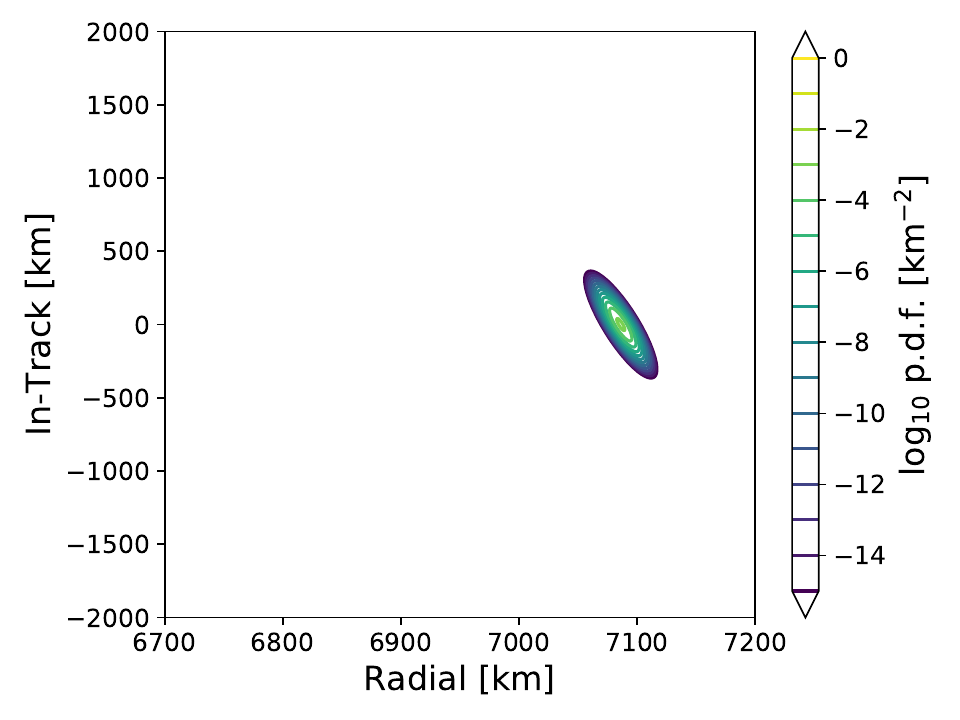}}\subcaptionbox{Silverman's regularization}{\includegraphics[trim= 0mm 0mm 29mm 0mm, clip,height=.4\linewidth]{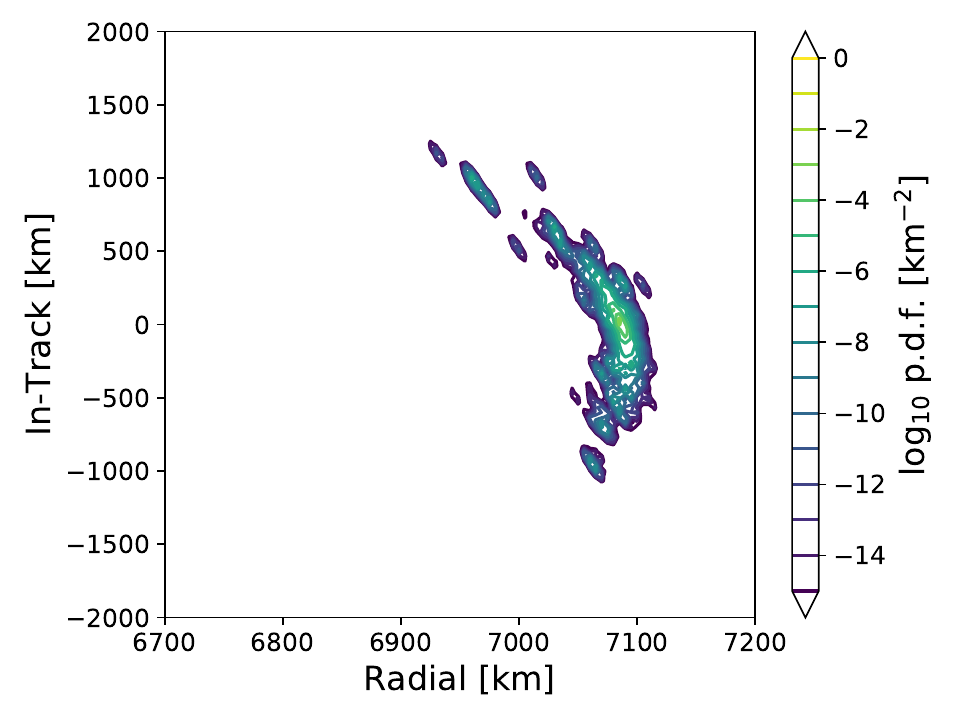}}
    \includegraphics[trim= 133mm 0mm 0mm 0mm, clip,height=.4\linewidth]{figparts_NT250000_NN1000_nx121.pdf} \\
    \subcaptionbox{Standard $k$-nearest neighbor}{\includegraphics[trim= 0mm 0mm 29mm 0mm, clip,height=.4\linewidth]{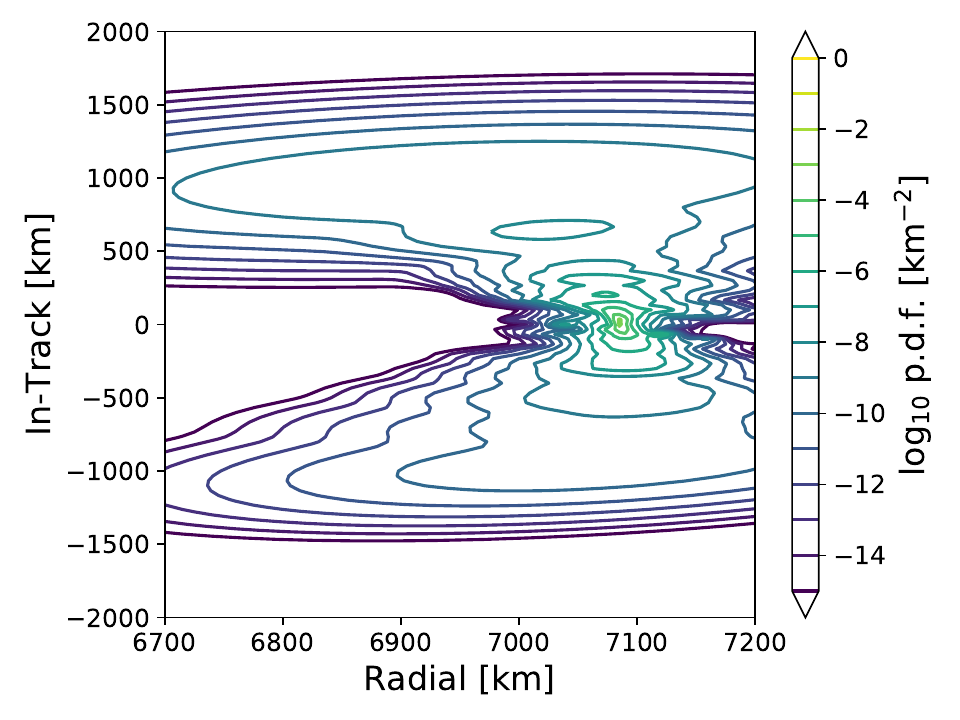}}
    \subcaptionbox{Proposed method}{\includegraphics[trim= 0mm 0mm 29mm 0mm, clip,height=.4\linewidth]{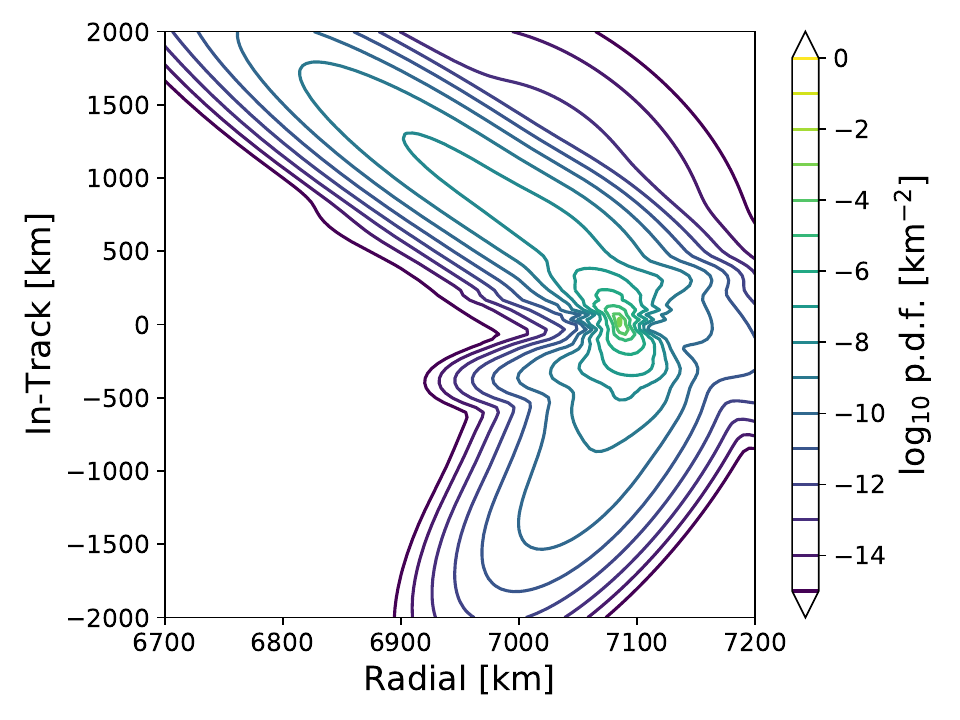}}
    \includegraphics[trim= 133mm 0mm 0mm 0mm, clip,height=.4\linewidth]{figparts_NT250000_NN1000_nx121.pdf}
    \caption{Transitional prior obtained with different regularization techniques.}
    \label{fig:regularization}
\end{figure*}

Figure~\ref{fig:regularization} shows the distributions obtained sampling according to Eq.~\eqref{eq:samplingMixture}, with different regularization techniques. Again, the distributions are obtained by propagating 2,500 particles, and, where needed, 99 are used for regularization. Note that while the proposed regularization approach appears more diffuse than the reference displayed in the right plot of Fig.~\ref{fig:transprior}, the total standard deviation does not change much before and after regularization, as evidenced in the last two rows of Table~\ref{table:stdevIS}.
The proposed approach provides tails orders of magnitude larger than both Silverman's regularization and the single Gaussian approach, while at the same time having a comparatively contained overall variance. Moreover, Silverman's approach leaves wide empty spaces; the voids are larger when looking at the actual 6 dimensions. Larger kernels could be used for Silverman's approach, but that would lead to larger total uncertainty. Further, a standard $k$-NN regularization approach has been tested. While in this case the tails are wider, the shape is not appropriately represented. The figure shows how importance sampling alone is not sufficient, and adaptive kernel density estimation is a necessary tool for distributions with super-Gaussian tails. While eventually the tail decay does become sub-Gaussian with the use of any Gaussian kernels, the synergy between adaptive kernel density estimation and importance sampling leads to non sub-Gaussian tails for a large portion of the domain. 

\subsection{Filter Update}
\label{sec:method}
The proposed filter update combines properties of the Gaussian Sum Filters (GSFs) and of iBLS filters. The \ac{gmm} obtained as in Sec.~\ref{sec:transprior} is updated using iBLS. All batches share the same multi-fidelity interpolation, and the weight of each component is updated based on its likelihood over the entire measurement pass. The batch is done over an augmented state that includes a constant acceleration over the pass. As the resulting uncertainty for each batch is Gaussian, marginalization to obtain joint position and velocity distributions of the target is straightforward.

Let the augmented state be 
\begin{equation}
\nonumber \bm{\chi}^i=\begin{bmatrix}
    {\bm{r}^i}^T &
    {\bm{v}^i}^T &
    {\bm{t}^i_p}^T,
\end{bmatrix}^T
\end{equation}
where $\bm{t}^i_p$ is the thrust.
The corresponding covariance is
\begin{equation}
    \nonumber P^i = \begin{bmatrix}
    P^i_{rr} & P^i_{rv} & 0_{3\times 3}\\
    P^i_{rv} & P^i_{vv} & 0_{3\times 3}\\
    0_{3\times 3} & 0_{3\times 3} & Q^i_p\\  
\end{bmatrix},
\end{equation}
where the thrust is given a prior uncertainty that is a function of the particle:
\begin{equation}
\label{eq:priorAcc}
    Q_p^i = \mathbb{I}_{3\times 3}\frac{\bm{t_i}^T\bm{t_i}}{3n_{\Delta V}}.
\end{equation}
The larger the thrust magnitude of the particle during the measurement gap, the larger the thrust uncertainty is through the measurement pass.
For a problem where $\bm{\chi}\sim \mathcal{N}\left(\bar{\bm{\chi}},P\right)$, $\bm{y}=\bm{h}\left(\bm{\chi}\right)+\bm{\epsilon}$, and $\bm{\epsilon}\sim\mathcal{N}\left(0,R\right)$, the MMSE estimate can be approximated by the EKF (\textit{e.g.},~\cite{barshalomnonlinear}):
\begin{align}
\label{eq:kf0}
 K_k^i &= P^i_{k+1|k}\left(H^i_{k}\right)^T \left(H^i_{k}P^i_{k+1|k}\left(H^i_{k}\right)^T+R_k\right)^{-1}\\
\label{eq:kf1}
    \bm{\chi}^i_{k+1|k+1}& = \bm{\chi}^i_{k+1|k} + K^i_k\left(\bm{h}\left(\bm{\chi}^i_{k+1|k}\right)-\mathbf{y}_{k+1}\right),\\
    \label{eq:kf2}
    P^i_{k+1|k+1} &= \left(I-K^i_k H^i_k\right)P_{k+1|k}^i,
\end{align}
where
\begin{equation}
    H^i_{k+1|k}=\left|\frac{\partial \bm{h}\left(\bm{\chi}\right)}{\partial \bm{\chi}}\right|_{\bm{\chi}=\bm{\chi}^i_{k+1|k}}.
\end{equation}

A GSF update consists of several EKF updates in parallel. Their means and covariances are updated as in Eqs.~\eqref{eq:kf0}-\eqref{eq:kf2} and the relative weight of each filter is also updated:
\begin{align}
    \hat{w}_{k|k}^i &= \frac{w^i_{k|k-1}}{\sqrt{{\left| 2\pi (HPH^T+R)\right|}}} e^{-\frac{1}{2}\left(\bm{h}\left(\bm{\chi}^i_{k+1|k}\right)-\mathbf{y}_{k+1}\right)^T\left(HPH^T+R\right)^{-1}\left(\bm{h}\left(\bm{\chi}^i_{k+1|k}\right)-\mathbf{y}_{k+1}\right)} \\
    {w}_{k|k}^i &= \frac{\hat{w}_{k|k}^i}{\sum_{j=1}^N\hat{w}_{k|k}^j},
\end{align}
where $w^i_{k|k-1}$ is the prior weight obtained by importance sampling.

An EnGMF usually processes measurements one at a time. Here, since multiple measurements are received within a short time frame, batch processors are used instead, one for each component. To mitigate the effects of the nonlinearities, the batch is iterated. The weights are computed only at the first pass, as if the batch were not iterated, so that components with smaller weights can be pruned before iterating. The weights are obtained by propagating the distribution uncertainty linearly during the pass, and computing the total likelihood of the measurement pass. For component~$i$, the posterior weight after processing the batch of observations is 
\begin{equation}
\label{eq:postWeights}
    \hat{w}^i_{k|k} =w^i_{k|k-1} \prod_{j=0}^{n_{p,k}} \frac{e^{-\frac{1}{2}\left(\bar{\mathbf{y}}^i_{k_j}-\mathbf{y}_{k_j}\right)\left(H^i_{k_j}P^i_{k_j|k-1}\left(H^i_{k_j}\right)^T+R_{k_j}\right)^{-1}\left(\bar{\mathbf{y}}^i_{k_j}-\mathbf{y}_{k_j}\right)^T }}{\sqrt{\left|2\pi\left(H^i_{k_j} P^i_{k_j|k-1}\left(H^i_{k_j}\right)^T+R_{k_j}\right)\right|}},
\end{equation}
where $n_{p,k}$ is the number of measurements in the $k$\textsuperscript{th} pass, the subscript $k_j$ stands for the $j$\textsuperscript{th} measurement of the $k$\textsuperscript{th} pass,
$\bar{\bm{y}}^i_{k_j}=\bm{h}\left(\bm{\chi}^i_{k_j|k-1}\right)$ is the expected measurement for component $i$ at measurement $k_j$, and
\begin{equation}
    P^i_{k_j|k-1} = F^i_{k_j,k_0} P^i_{k_0|k-1}\left(F^i_{k_j,k_0}\right)^T,
\end{equation}
where $F^i_{k_j,k_0}$ is the state transition matrix of component $i$ from the start of the pass to $k_j$. While multi-fidelity state transition matrix propagation has been developed~\cite{zucchelli2023hmc}, finite difference methods have been used here.

Any component whose posterior weight is less than 10\textsuperscript{-10} times the maximum weight is pruned after the first iteration.
After the lower likelihood components are pruned, the update of mean and variance of the mixands can be performed. Also in this phase, multi-fidelity propagation is used. 

The process is iterated in order to compute the partials $F$ and $H$ as closely as possible to the actual estimate. Even having the exact partials over the estimate is in theory not sufficient to remove bias from the filter. However, the fact that those partials are exact where the probability density is highest improves the estimate.

Two multi-fidelity interpolations are computed, both along the entirety of the pass: the first one is computed for the full Gaussian mixture, and is used for the first iteration only. The second multi-fidelity interpolation is computed for the second iteration, using only the non-pruned components, and is kept for all subsequent iterations. After the first iteration, two things happen concurrently: 1) many components are pruned, and therefore the interpolation domain is reduced, and 2), the majority of the update for all other components is already applied, driving the interpolation domain towards to the more important region of the state space. These two facts help improve the approximation. While additional multi-fidelity interpolations may improve the propagation, every different interpolation has different errors, therefore causing the iterative process to oscillate.

\subsection{Filter Summary}
\begin{figure*}[htb]
    \centering
     \includegraphics[width=0.9\linewidth]{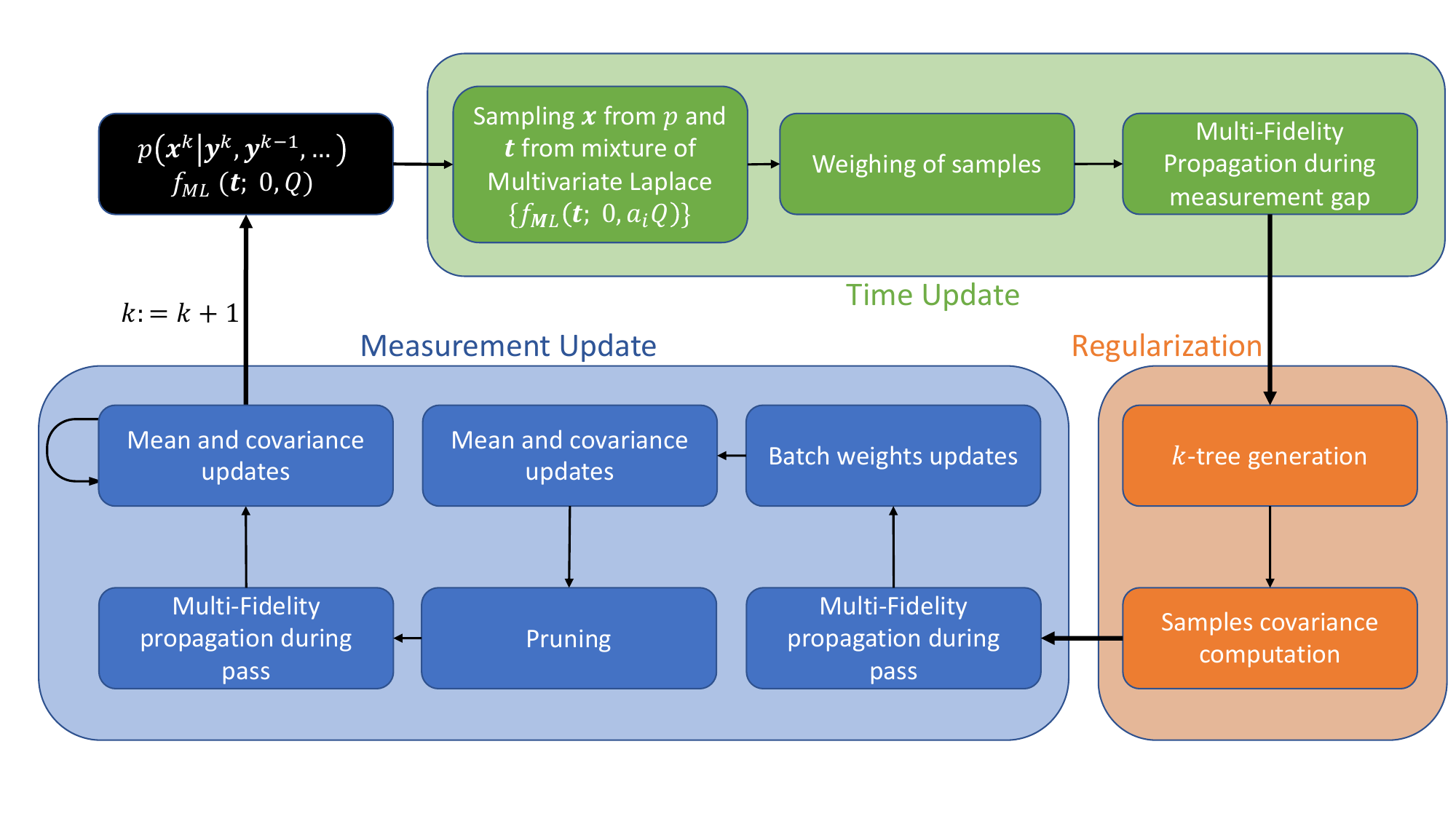}
    \caption{Overall architecture of the filter.}
    \label{fig:scheme}
\end{figure*}
This section summarizes the schematics of the filter, depicted in Fig.~\ref{fig:scheme}.
At the end of a measurement pass, random target states $[{\bm{x}^i}^T,{\bm{v}^i}^T]^T$, $i=1,...,N$ are sampled from an initial distribution. Each sample is associated to a thrust profile $\bm{t}^i_k$, sampled from a sampling distribution that is the mixture described in equation~\eqref{eq:samplingMixture}, and are given the prior weight $w_i$ as in Eq.~\eqref{eq:priorWeights}. The samples are propagated with the low-fidelity model until the next measurement. Out of those, important samples are selected according to the multi-fidelity stochastic collocation algorithm, and are propagated with high-fidelity model. All low-fidelity samples are corrected with the multi-fidelity stochastic collocation method. Then, a 6D-tree is generated for efficient $k$-NN search. The tree is generated using states transformed into canonical units. Every particle is transformed into a Gaussian distribution, with covariances computed according to Eqs.~\eqref{eq:gating}-\eqref{eq:ensemblecovariance}. Each mean and covariance is propagated with state transition matrices for the duration of the pass, using again the stochastic collocation multi-fidelity approach. Expected measurements and measurement Jacobians are computed for every component at each measurement of the pass. The weights of each component are computed according to Eq.~\eqref{eq:postWeights}, and normalized. Components with relative weights smaller than a given threshold are pruned. All components are updated for one round of batch least squares. After the first update, a new stochastic collocation multi-fidelity interpolation is generated over the pass, and all remaining components are iteratively updated with a batch. The batch is performed over an augmented state that includes a constant acceleration. The prior variance of the acceleration is computed according to Eq.~\eqref{eq:priorAcc}. The second multi-fidelity stochastic collocation interpolation is then held constant during the remaining interpolation of the batch least squares. Finally, one obtains the updated augmented state and covariance of each component at the beginning of the measurement pass, together with a constant (in the RIC frame) thrust vector, which need to be propagated to the final epoch of the pass. At this point one has a posterior distribution at the last measurement of a pass, and can proceed forward repeating the algorithm for the next measurement gap.

\section{Results}
\label{sec:results}
\begin{table}
\centering
\caption{Orbit state initial conditions and $1\sigma$ uncertainty}\label{table:initial_conditions}
	\begin{tabular}{ccccccc}
		\hline
		   & ${a}$ [km] & ${e}$ [-] &  ${i}$ [deg] &  ${\Omega}$ [deg] &  ${\omega}$ [deg] &  ${\nu}$ [deg]  \\
		& 7,078 & 7.7$\times10^{-4}$ & 45 & 0 & 90 & 0 \\\hline
		  &   &  &  &  &  &   \\
		  &  ${x}$ [km] & ${y}$ [km] &  ${z}$ [km] &  ${\dot{x}}$ [m/s] &  ${\dot{y}}$ [m/s] &  ${\dot{z}}$ [m/s]  \\
           & 0.0 & 5,001.048 & 5,001.048 & -7,510.139 & 0.0 & 0.0 \\
		$\sigma$ & 1.0 & 1.0 & 1.0 & 1.0 & 1.0 & 1.0\\
  \hline
	\end{tabular}
\end{table}
The proposed algorithm has been tested on several scenarios. All scenarios share the same initial conditions, same true dynamical models, same filter propagation dynamical models, same measurement models, and same ground station network. The only differences consist of the thrust profiles and the observation passes frequency. 
The initial conditions are distributed as in Table \ref{table:initial_conditions}. The model used for the truth is detailed in Table \ref{table:force_models_fullsim} with additional parameters in Table~\ref{table:force_model_params}. 
The settings for the filters, such as number of samples used, number of neighbors used for each kernel, and number of distributions used in the proposal density mixture, are also all the same, unless otherwise specified, except for the standard deviation of the thrust.
Since in reality one never has a propagation model that is the same as the true physics, the high-fidelity model used in this work neglects drag and solar radiation pressure~(SRP). For context, these accelerations in the tested cases are no larger than 0.5~\textmu m/s\textsuperscript{2}. There is no additional process noise in the filter to account specifically for model mismatch, as the resulting uncertainty is generally smaller than the uncertainty introduced by the maneuvers. The low- and high-fidelity models used for propagation are summarized in Table~\ref{table:force_models_fullsim}; details can be found in Refs.~\cite{folkner_2014,pavlis_2012,petit_2010}.

\begin{table}
\centering
	\caption{Force models for low- and high-fidelity propagators, and for the truth.}\label{table:force_models_fullsim}
	\begin{tabular}{c|ccc}
		\hline
		 Model &  Low-Fidelity &  High-Fidelity & Truth \\
		\hline
		Central Body Gravity & Two-Body and $J_2$ & 70$\times$70\cite{folkner_2014} & 70$\times$70\cite{folkner_2014} \\
		Third-Body Perturbations & None & Sun and Moon\cite{pavlis_2012}  & Sun and Moon\cite{pavlis_2012} \\
		Solar Radiation Pressure & None & None  & Cannonball \\
		Atmospheric Drag & None & None  & Cannonball \\
		Coordinate System Reduction & None & IAU2006 \cite{petit_2010} & IAU2006 \cite{petit_2010}\\
  \hline
	\end{tabular}
\end{table}
Three ground stations generate radar measurements.
One ground station is located at $44.77^\circ$ latitude, $83.65^\circ$ longitude, and at a distance from Earth's center of $R_e$. The second and third stations have the same longitude and radius, but latitudes of $0^\circ$ and $-44.77^\circ$, respectively. 
Radar measurements are generated every 60~s when two conditions are met: 1) the spacecraft is visible from one of three stations, and 2) when the time from the latest pass is larger than a specified value, dependent on the scenario. The latter condition is included to force the filter to work under conditions of sparse data.
During the $k$\textsuperscript{th} pass, a set of measurements $Y_k=\left[\mathbf{y}_{k_0},...,\mathbf{y}_{k_{n_{p,k}}}\right]$ is obtained, where $n_{p,k}$ is the number of measurements during pass $k$, and $\mathbf{y}_{k_i}=\left[r_{k_i},\dot{r}_{k_i}, \alpha_{k_i},\delta_{k_i}\right]^T$, where $r$ is the range, $\dot{r}$ is the range-rate, $\alpha$ and $\delta$ are the topocentric right ascension and declination of the position, respectively. The measurement variance $R$ is
\begin{equation}
 \nonumber   R =\text{diag}\left(\left[(0.1\, \text{km})^2, \quad(0.003\, \text{km/s})^2, \quad(0.15\, \text{deg})^2,\quad (0.15\, \text{deg})^2\right]\right),
\end{equation}
and measurement errors are uncorrelated in time.

Except for Sec.~\ref{sub:versparse}, all tested filters use 2,500~propagated samples, and perform regularization with 99~neighbors. The samples are drawn from a mixture of nine multivariate Laplace distributions with diagonal variance; their standard deviations are equally spaced in logarithmic scale. The minimum standard deviation used is equal to the filter's set 1$\sigma$ standard deviation, and the maximum is set to 800~\textmu m/s\textsuperscript{2}. The distribution with smallest uncertainty has weight~0.3, and all others have equal weight. The iterative batch least squares performs ten iterations; at each iteration, the estimate is updated by a vector that is the computed updated, multiplied by 0.7.

Four scenarios are tested. Every scenario may include different cases; each case is tested over a Monte Carlo of fifty trajectories. The first scenario, in Sec.~\ref{sub:consistency}, is a consistency check.
The scenario in Sec.~\ref{sub:constthrust} is a constant in-track thrust case, during which the target spirals outwards.
Tests include different values of the thrust and different data gap durations.
The results are compared against another filter from literature.
The third scenario, described in Sec.~\ref{sub:aporaise},
is an apogee raising maneuver, which has an intermittent in-track thrust profile, causing the satellite to be harder to track.
This scenario is tested over different values of the target's thrust and the filter's assumed thrust, to prove robustness over the filter's parameter.
The same scenario is tested in a maneuver initiation case, to show the capability of the approach to react to the beginning of a maneuvering phase, and in cases with even sparser observation passes,
as few as one every 16~hours, to stretch the filter's capabilities.
Finally, in Sec.~\ref{sub:autocorr}, the filter is tested for cases where the thrust is directed in all three directions,
and changes almost continuously.

\begin{table}
\centering
	\caption{Force model and satellite parameters for all cases.}\label{table:force_model_params}
	\begin{tabular}{cc}
		\hline
		 Parameter &  Value \\
		\hline
		Satellite Mass & 500 kg \\
		Satellite Drag and SRP Area & 1 m$^2$ \\
		$C_D$ & 2.0 \\
		$C_R$ & 1.5 \\
		Epoch Time & 2455200.5 UTC \\
		Earth Radius $R_\oplus$ & 6378.1363 km \\
		Gravitation Parameter & 398600.4415 \\
  \hline
	\end{tabular}
\end{table}

\subsection{Consistency check: random thrust}
\label{sub:consistency}
This case is to be considered a toy problem and is used to evaluate the consistency of the proposed filter when its assumptions are met. To this end, the true thrust is generated in a way that is as similar as possible to how the filter samples the thrust profiles. Note that for all other scenarios the filter cannot be consistent. That is because in those scenarios all trajectories in a same Monte Carlo trial share the same thrust profile.

\begin{figure*}[htb]
    \centering
     \includegraphics[trim= 0mm 0mm 0mm 0mm, clip,width=3.25in]{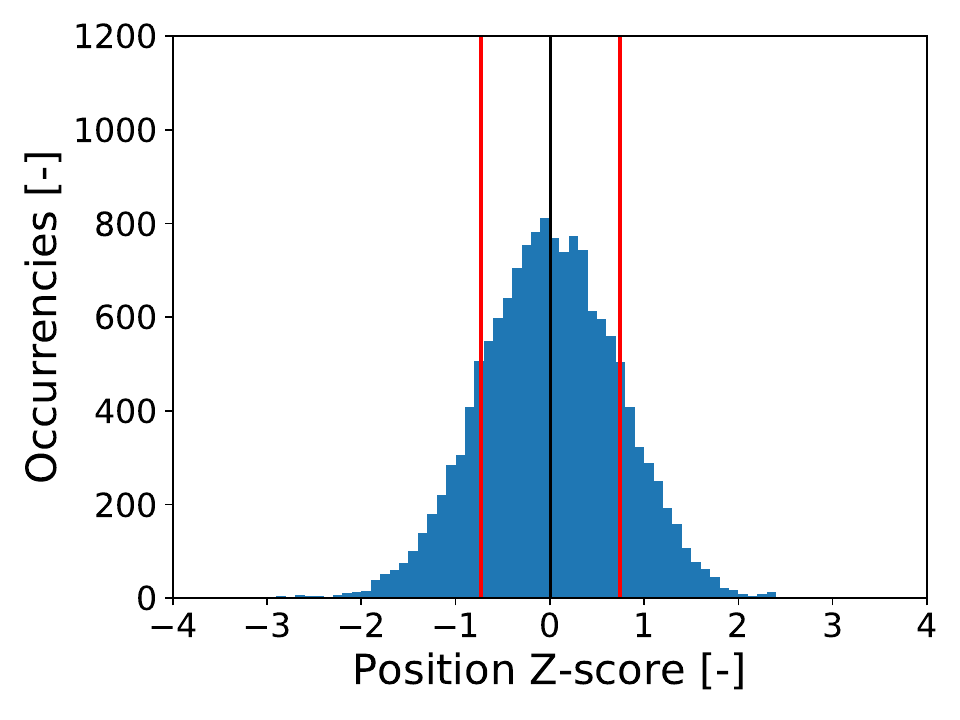}\\
     \includegraphics[trim= 0mm 0mm 0mm 0mm, clip,width=3.25in]{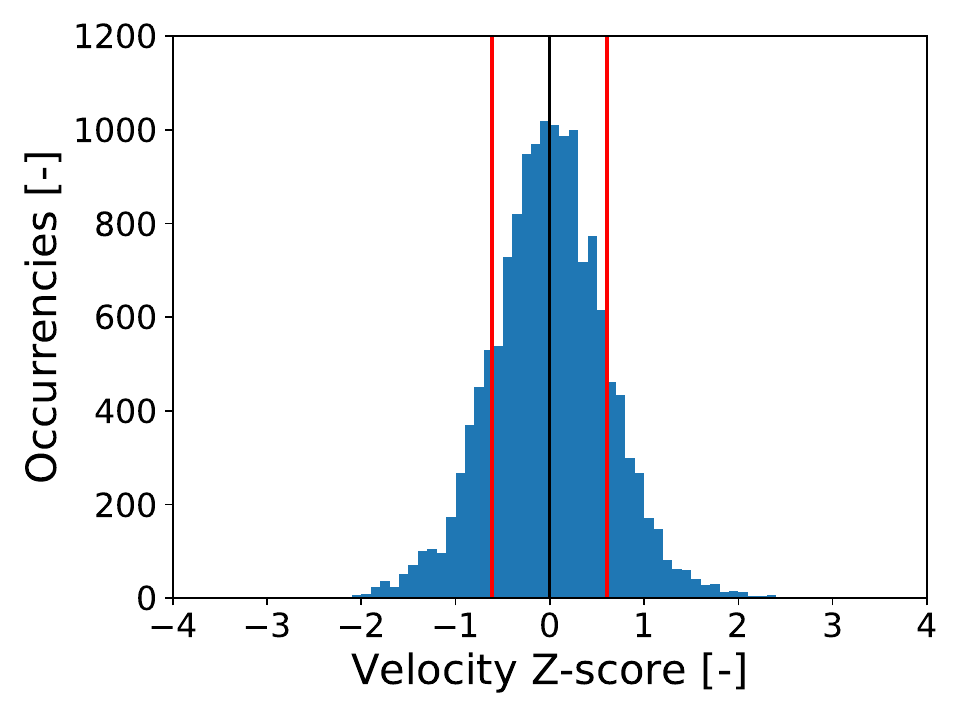}
    \caption{Z-score histograms for Sec.~\ref{sub:consistency}, and corresponding statistical means and standard deviations.}
    \label{fig:hist5e8}
\end{figure*}

The actual thrust and the filter's thrust samples are drawn from the same distribution, a three-dimensional multivariate Laplace with standard deviation equal to 50~\textmu m/s\textsuperscript{2} per direction. The true thrust is sampled every 16,800~s, the approximate average time between each measurement pass, and is  constant for that duration. Accordingly, only for this scenario and the next, the filter assumes constant thrust between every measurement pass, but magnitude and directions are unknown. Figure~\ref{fig:hist5e8} displays histograms of the Z-scores of the estimates, the ratios between actual error and filter's estimated uncertainty. The statistical mean is the black vertical line and the 1$\sigma$ statistical standard deviation is marked by the red lines. For a perfectly consistent filter, the latter should be equal to 1. The filter's standard deviation of the error is slightly conservative, as expected, since the regularization inflates the uncertainty of the transitional prior. Most importantly, since there are no large systematic differences between the physics model and the filter's model, the distributions have biases much smaller than their standard deviation. The \ac{rmse} of the position is 507~m and the \ac{rmse} of the velocity is 0.978~m/s. Fewer than 0.1\% of estimates fall outside of the 3$\sigma$ bounds.
 
\subsection{Constant Thrust}
\label{sub:constthrust}

In this more realistic scenario, the thrust is constant during the entire simulation, and directed in-track: the proposed filter does not know the magnitude of the thrust, nor its direction, and assumes constant thrust within every measurement gap, like in Sec.~\ref{sub:consistency}. The latter assumption is dropped in the next subsection.
Only for this scenario, the filter is compared against the variable-state dimension filter~(VSDF) proposed by Goff et al.~\cite{goff2015orbit}. To reduce complexity, the single model version of the VSDF is used; the maneuver standard deviation for the VSDF is set equal to the magnitude of the true acceleration. The proposed filter here assumes a thrust of 1~\textmu m/s\textsuperscript{2}. This gives the VSDF an advantage based solely on tuning parameters. Note that a second advantage for the VSDF comes from the fact that in this scenario the true thrust is constant; while the proposed filter can handle varying thrust, the VSDF cannot.

\begin{figure*}[htb]
    \centering
     \includegraphics[trim = 18mm 60mm 22mm 70mm,clip,width=.5\linewidth]{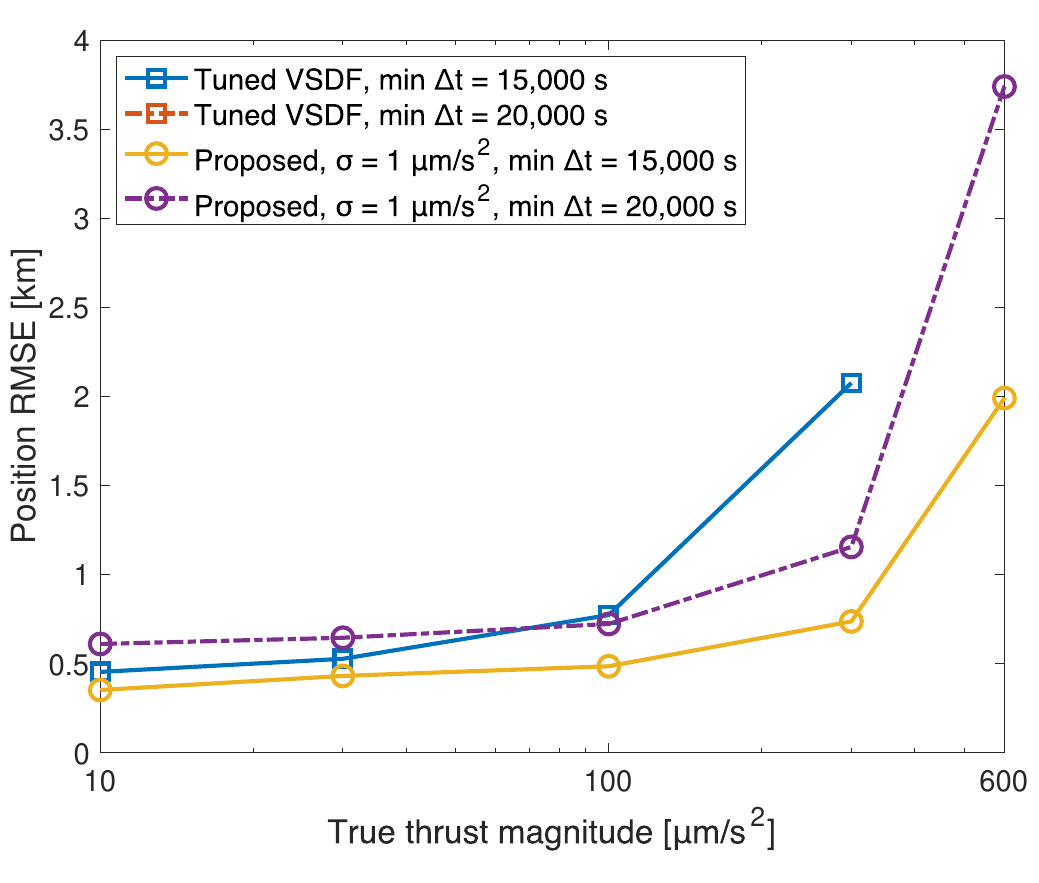}\includegraphics[trim = 18mm 60mm 22mm 70mm,clip,width=.5\linewidth]{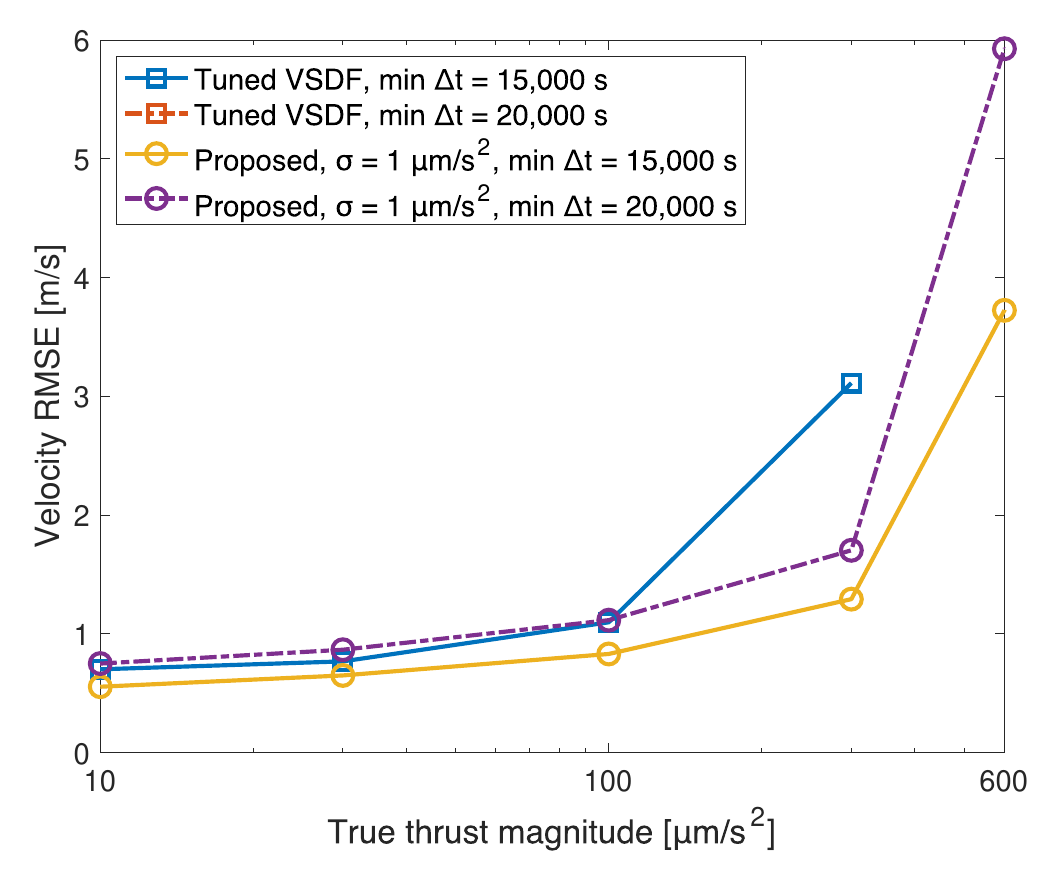}
    \caption{RMSE for the filters applied to the scenario of Sec.~\ref{sub:constthrust}.}
    \label{fig:resultsscenarioconst}
\end{figure*}
\begin{figure*}[htb]
    \centering     
     \includegraphics[trim= 0mm 0mm 0mm 0mm, clip,width=3.25in]{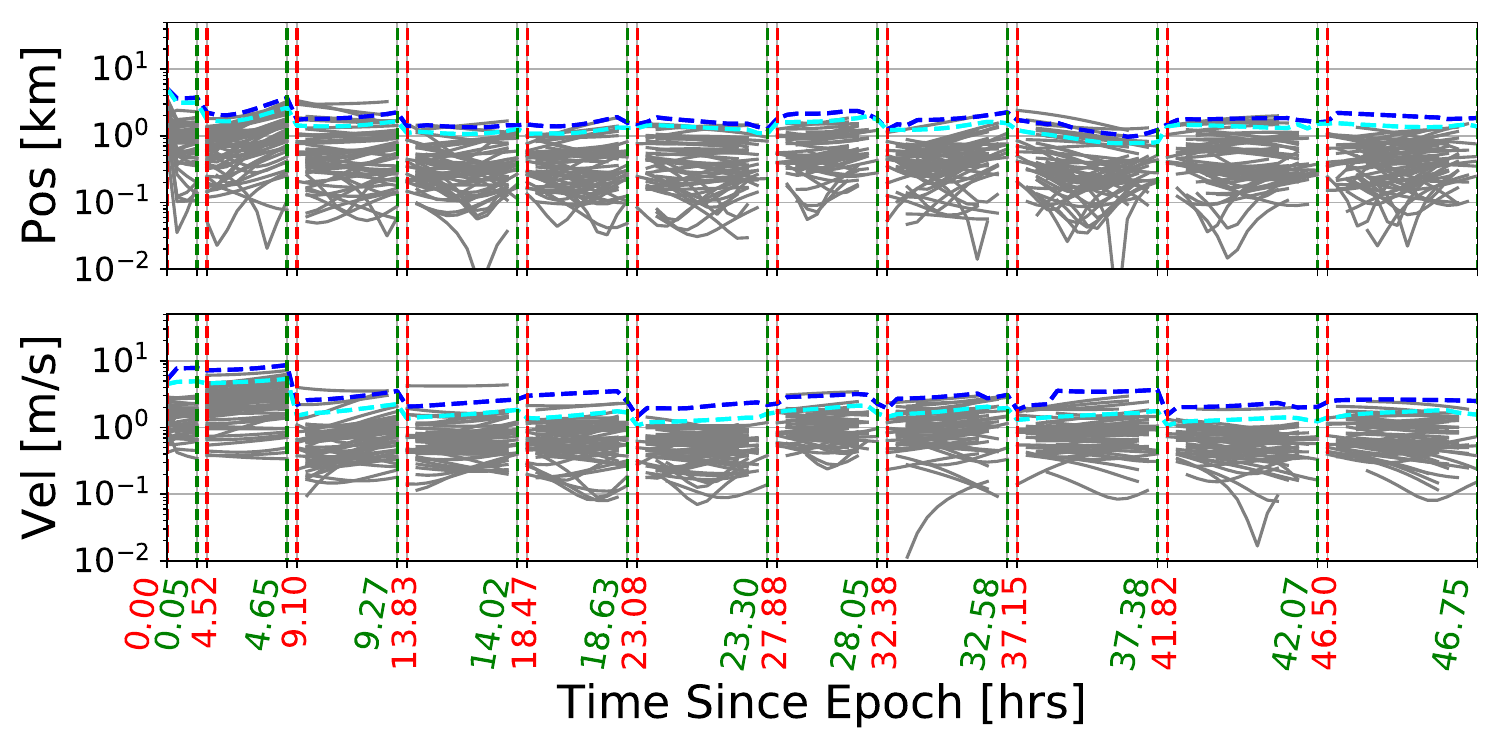}
    \caption{Errors for the proposed filter, true thrust constant at 300~\textmu m/s\textsuperscript{2}.}
    \label{fig:MC11}
\end{figure*}

The filters have been tested for a variety of true thrust magnitudes and for cases with minimum measurement gaps of 15,000~s and 20,000~s as shown in Fig.~\ref{fig:resultsscenarioconst}. Each data point is obtained from a Monte Carlo trial of 50 trajectories, and every simulation lasts 48~hours. Because of the fact that the VSDF linearizes models to approximate the distributions, it has at least one diverging trajectory for any of the cases with 20,000~s measurement gaps, as well as the case with 15,000~s and 600~\textmu m/s\textsuperscript{2} acceleration. Therefore, those results are not reported. On the other hand, the proposed filter can track the target in all cases, albeit some signs of divergence occur when the true thrust is 600~\textmu m/s\textsuperscript{2} and the filter assumes 1~\textmu m/s\textsuperscript{2} acceleration. Further, for the cases in which the VSDF does not diverge, it is always outperformed by the proposed filter.
Figure~\ref{fig:MC11} shows the errors estimates of the proposed filter when the true thrust is 300~\textmu m/s\textsuperscript{2}. The plot skips over the measurement gaps; measurement passes start and end at, respectively, red and green dashed vertical lines. The blue and cyan dashed lines are, respectively, the maximum and average (over the 50 Monte Carlo runs) 3$\sigma$ uncertainties expected by the filter at the given time stamp. The plot shows that the filter successfully tracks the maneuvering object, and almost all estimates stay within the 3$\sigma$ bounds.

\subsection{Apogee Raising Maneuver}
\label{sub:aporaise}

\begin{figure*}[htb]
    \centering
     \includegraphics[width=3.25in]{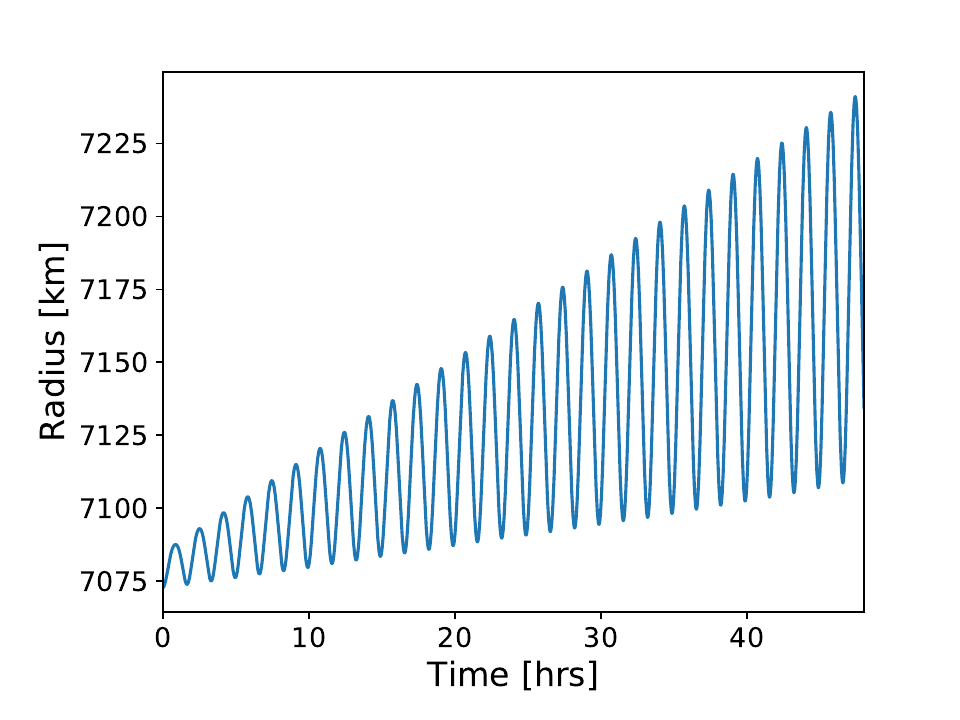}
    \caption{Example radius of one of the trajectories from Sec.~\ref{sub:aporaise}.}
    \label{fig:orbit_scenario2}
\end{figure*}

In this third scenario the spacecraft performs an apogee raising maneuver. The intermittent, in-track thrust is on and constant for half of an orbit, and off for the remaining half. An example of the orbital radius of the true trajectory is shown in Fig.~\ref{fig:orbit_scenario2}.
Because of the intermittency of the thrust, combined with the physics of the problem, filters that assume constant thrust during a measurement gap (like the ones from Sec.~\ref{sub:constthrust}) perform very poorly when applied to this problem. For example, with a true acceleration of 300~\textmu m/s\textsuperscript{2} and an assumed acceleration of 25~\textmu m/s\textsuperscript{2}, a proposed filter that assumes constant thrust produces more than one third of the estimates exceeding the 3$\sigma$ bounds, and the 20$\sigma$ bounds are reached.
For this scenario, the number of thrust segments in a measurement gap is chosen as follows:
\begin{equation}
\label{eq:nDeltaV}
    n_{\Delta V} = \text{floor}\frac{4\Delta t}{2\pi}\sqrt{\frac{\mu}{a^3}},
\end{equation}
where $\Delta t$ is the time between the end of the previous measurement pass and the start of the next, and $a$ is the semi-major axis of the osculating orbit at the start of the measurement gap. Hence, there are approximately 4 segments per orbit.
\subsubsection{Base Apogee Raising Scenario}
\label{subsub:aporaise}
Direction, thrust profile, and thrust magnitude are unknown to the filter. The filter additionally assumes that the thrust is repeated every 4 thrust segments within the same measurement gap. This assumption is generally acceptable for relatively short measurement gaps, and it is dropped for cases with larger gaps as in Sec.~\ref{sub:versparse}. Like in all previous scenarios, there is no assumption of repeating the thrust profiles between different measurement gaps.

\begin{figure*}[htb]
    \centering
     \includegraphics[trim=0mm 0mm 0mm 0mm,clip,width=3.25in]{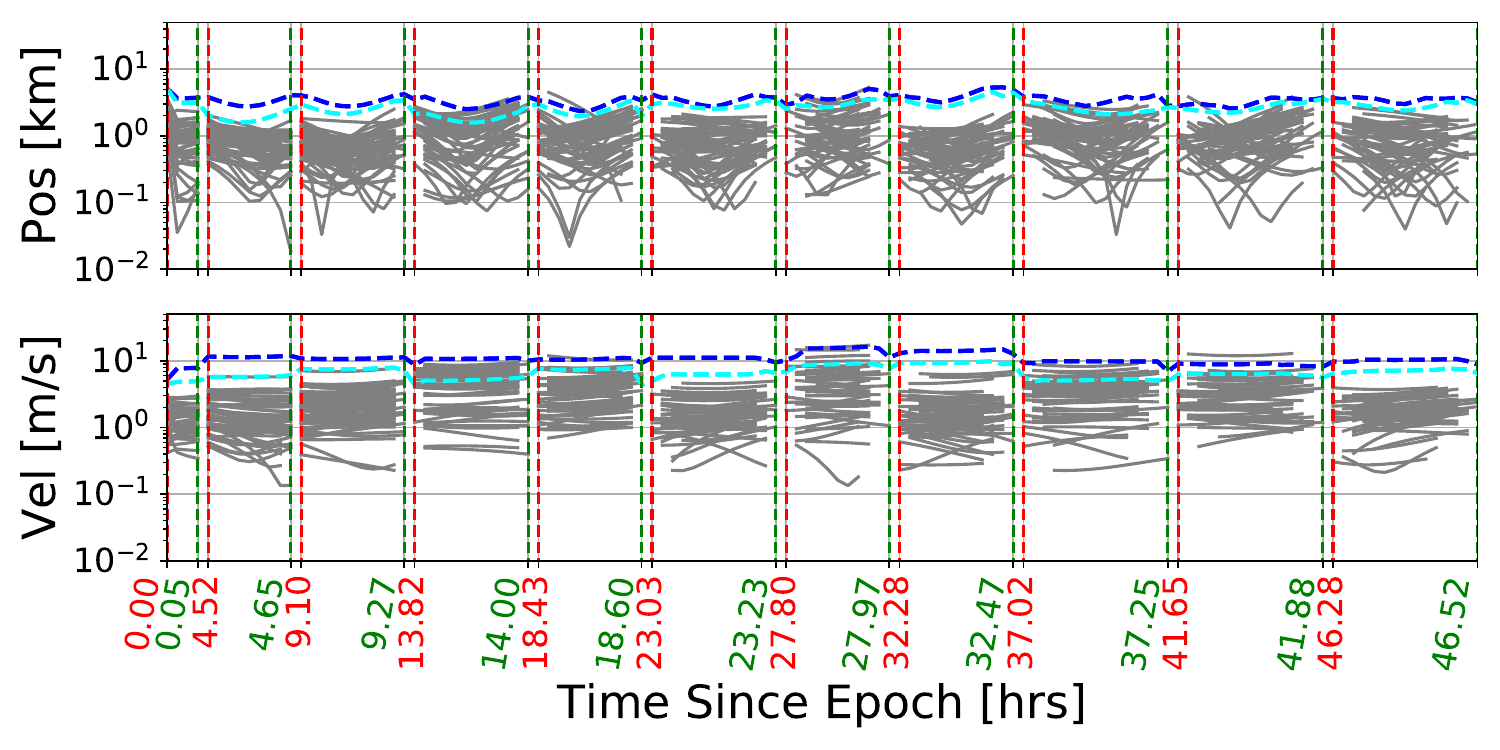}
    \caption{Estimate errors for Case B1 of the scenario from Sec.~\ref{sub:aporaise}.}
    \label{fig:MC2_24}
\end{figure*}

Figure~\ref{fig:MC2_24} shows the error and corresponding maximum and average filter 3$\sigma$ estimate uncertainty for filter's standard deviation of 1~\textmu m/s\textsuperscript{2},
and with true target acceleration of 300~\textmu m/s\textsuperscript{2}.
Comparing the plot with Fig.~\ref{fig:MC11} it is possible to obtain confirmation that this scenario is more challenging than the previous one. In fact, the \ac{rmse} is 52\% larger in the position and 169\% larger in the velocity. Further, the number of outliers is greatly increased, from a negligible number to more than 7\%.

Table~\ref{table:resultsIT} summarizes the performance of the filters with different settings and different true maneuvers.
The ratio between actual thrust and the filter's standard spans three orders of magnitude.
The general trend is that the filter is more robust, and more accurate, whenever the actual and expected thrust are most similar in magnitude.
For same filter settings, larger thrust leads to larger errors and larger uncertainties, together with a larger number of outlier estimates. For the highest thrust case, the two filters with standard deviation of 1~and 6.25~\textmu m/s\textsuperscript{2} have an amount of outliers that exceed the expected amount even for univariate Laplace distributions, which is 1.5\%;
however, that can be considered acceptable given the systematic error inherent to the scenario.
Moreover, as shown in Fig.~\ref{fig:MC2_24} even the most challenging case does not diverge.\
Last, given everything else equal, filters whose 1$\sigma$ uncertainty differs by two orders of magnitude provide estimates whose accuracy differs by less than 30\%, showing that the proposed filter is largely insensitive to the process noise tuning.

\begin{table}[htbp]
\caption{Filters performance over 50~Monte Carlo trials for  the scenario from Sec.~\ref{sub:aporaise}.}
\label{table:resultsIT}
\centering
\begin{tabular}{cccccc}
\hline
Case & True magnitude & Filter 1$\sigma$ & Pos. RMSE & Vel. RMSE & \% outside~3$\sigma$\\
\hline
A1 & 30~\textmu m/s\textsuperscript{2} &  1~\textmu m/s\textsuperscript{2} & 427~m & 0.773~m/s &   0.2\% \\
A2 & 30~\textmu m/s\textsuperscript{2} &  6.25~\textmu m/s\textsuperscript{2} & 420~m & 0.753~m/s &  \textless \, 0.1\% \\
A3 &30~\textmu m/s\textsuperscript{2} &  25~\textmu m/s\textsuperscript{2} & 414~m & 0.741~m/s &  \textless \, 0.1\% \\
A4 &30~\textmu m/s\textsuperscript{2} &  100~\textmu m/s\textsuperscript{2} & 524~m & 0.989~m/s &  \textless \, 0.1\% \\
B1 & 300~\textmu m/s\textsuperscript{2} &  1~\textmu m/s\textsuperscript{2} & 1,117~m & 3.472~m/s &  7.7\% \\
B2 & 300~\textmu m/s\textsuperscript{2} &  6.25~\textmu m/s\textsuperscript{2} & 1,047~m & 3.118~m/s &   4.9\% \\
B3 & 300~\textmu m/s\textsuperscript{2} &  25~\textmu m/s\textsuperscript{2} & 944~m & 2.678~m/s &  1.6\% \\
B4 &300~\textmu m/s\textsuperscript{2} &  100~\textmu m/s\textsuperscript{2} & 870~m & 2.483~m/s &  0.2\% \\
 
\hline
\end{tabular}
\end{table}

\subsubsection{Maneuver Initiation}
\label{sub:manInit}
\begin{figure*}[htb]
    \centering \includegraphics[width=3.25in]{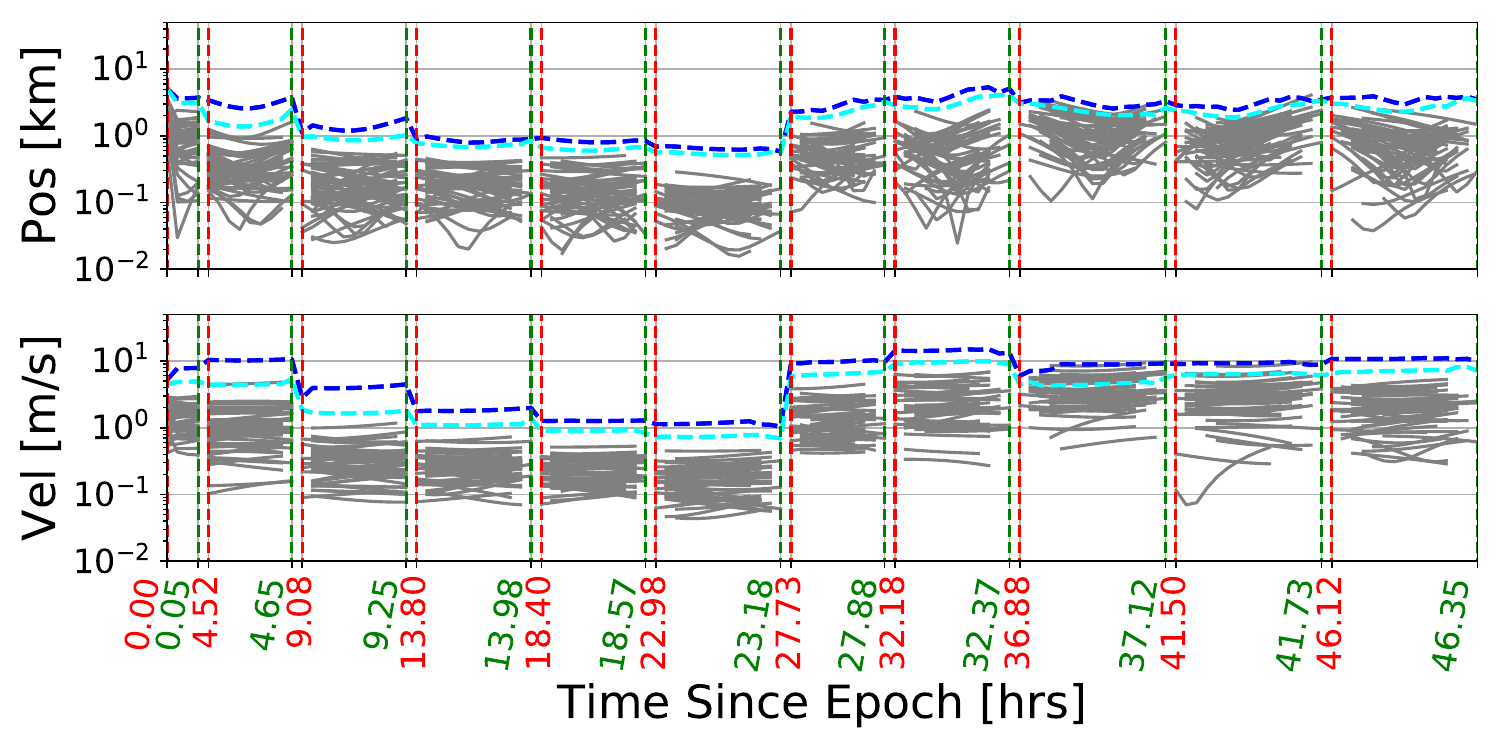}
    \caption{Estimate errors for the scenario of Sec.~\ref{sub:manInit}.}
    \label{fig:initiation} 
\end{figure*}

In this scenario the target flies ballistically for the first 24 hours and only then the apogee raising maneuver begins. The goal of this test is to show that the filter can also automatically switch between maneuvering and non maneuvering situations, and does not require any additional or external maneuver detection capabilities to operate.
Figure~\ref{fig:initiation} shows position and velocity errors for the extreme case with 1~\textmu m/s\textsuperscript{2} process noise standard deviation and 300~\textmu m/s\textsuperscript{2} actual acceleration. The uncertainty initially shrinks, because the steady-state covariance for the non-maneuvering problem is smaller than the initial uncertainty. Once the maneuver begins, the uncertainty increases accordingly.
 
\subsubsection{Very Sparse Observations}
\label{sub:versparse}
\begin{figure*}[htb]
    \centering
     \subcaptionbox{Data gap $\geq$ 12~hrs}{\includegraphics[trim=0mm 0mm 0mm 0mm,clip,width=3.2in]{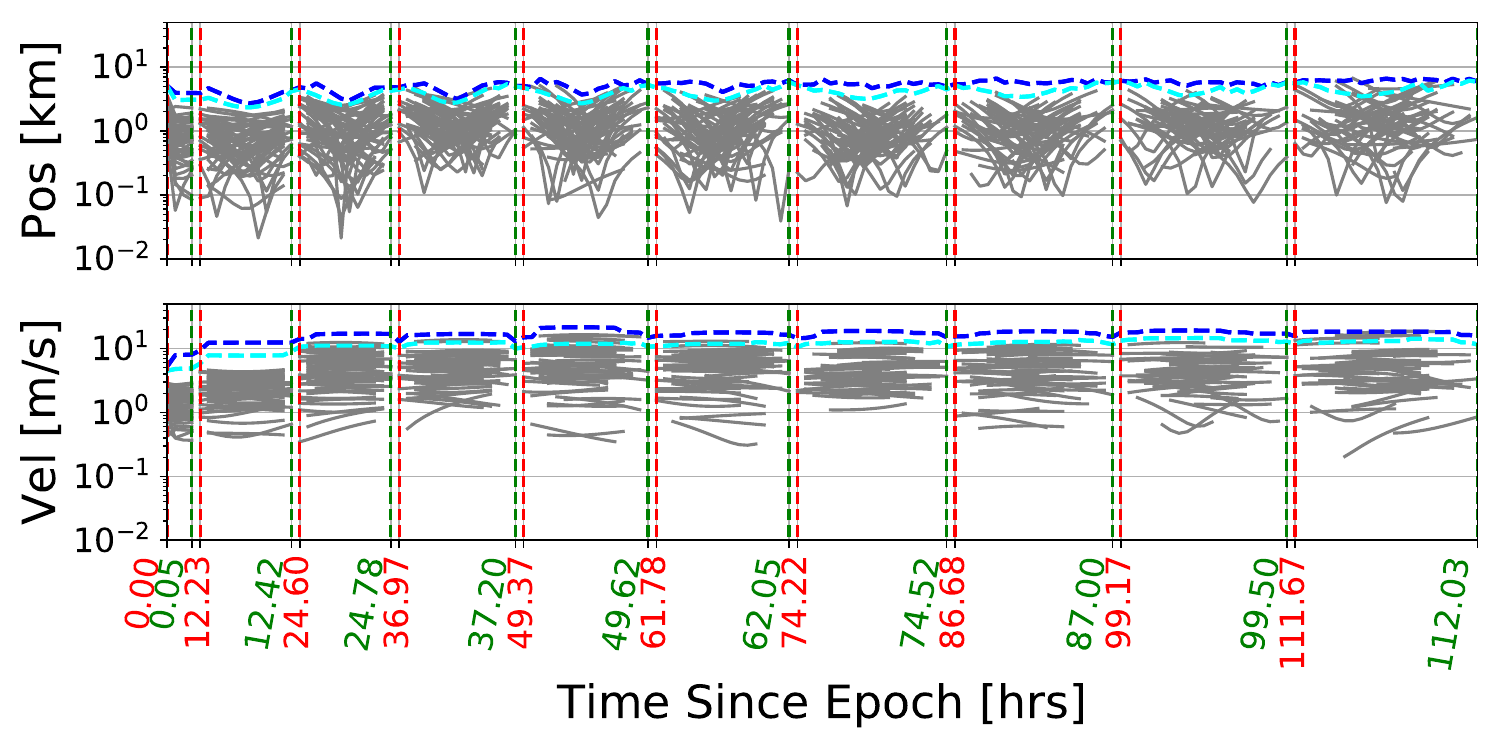}}
     \subcaptionbox{Data gap  $\geq$ 16~hrs}{\includegraphics[width=3.2in]{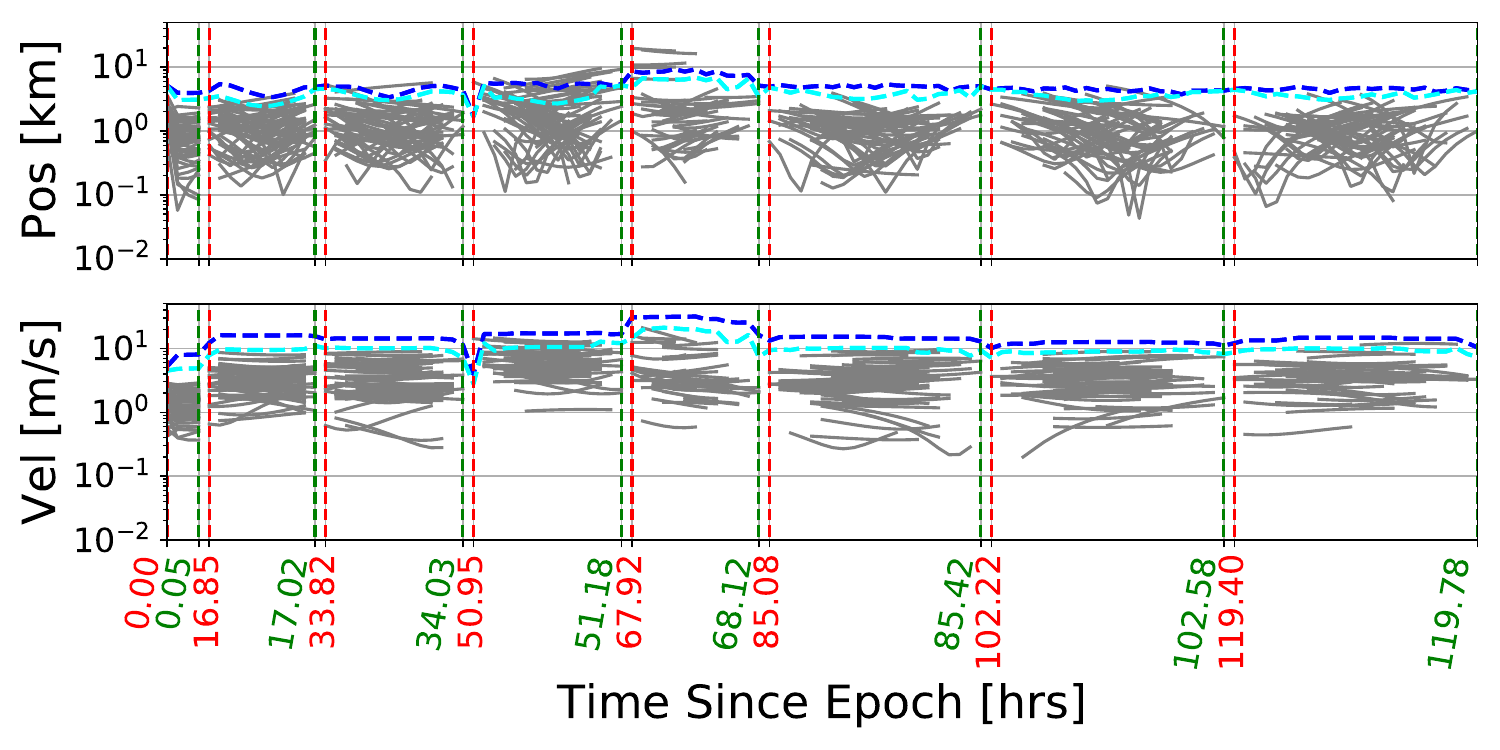}}
    \caption{Estimate errors for the scenario from Sec.~\ref{sub:versparse}}.
    \label{fig:versparse10k}
\end{figure*}

The filter's capability is further stretched with longer data gaps. The frequency of measurement passes is reduced such that gaps are larger than 12~hours in case~(a), and 16~hours in case~(b). The simulation lasts 5~days. As before, the true thrust is intermittent, and the filter assumes continuous segments of thrust of about one fourth of an orbit. Differently from before, now the thrust is not repeated between orbits, the filter's standard deviation is kept to 1~\textmu m/s\textsuperscript{2}, and the filter uses 10,000 particles, regularized using the 199 nearest neighbors.~\textmu m/s\textsuperscript{2}.
Figure~\ref{fig:versparse10k} shows the errors for a Monte Carlo trial tracking an object accelerating with thrust of 200~\textmu m/s\textsuperscript{2}. For case~(a) all trajectories are tracked successfully, with 2.1~\% of outliers estimates outside the 3$\sigma$ bounds. For case~(b), one out of 50 trajectories diverges, not reported in the plot; among the converging trajectories, the number of outliers outside 3$\sigma$ is 3.6\%; the position \ac{rmse} is 2.091~km and the velocity \ac{rmse} is 4.522~m/s. With same data gaps as case (a), but larger true thrust of 300~\textmu m/s\textsuperscript{2}, one out of 50 trajectories diverges after 10 days of simulation. The capabilities of the filter can generally be improved by increasing the number of particles, while sublinearly increasing the number of neighbors used for regularization. For example, when tracking case (a) with 2,500 particles, the outliers are 10\%, suggesting a trend of improving results with increasing sample number.

\subsection{Quickly Varying Thrust}
\label{sub:autocorr}
\begin{figure*}[htb]
    \centering
    \includegraphics[width=3.25in]{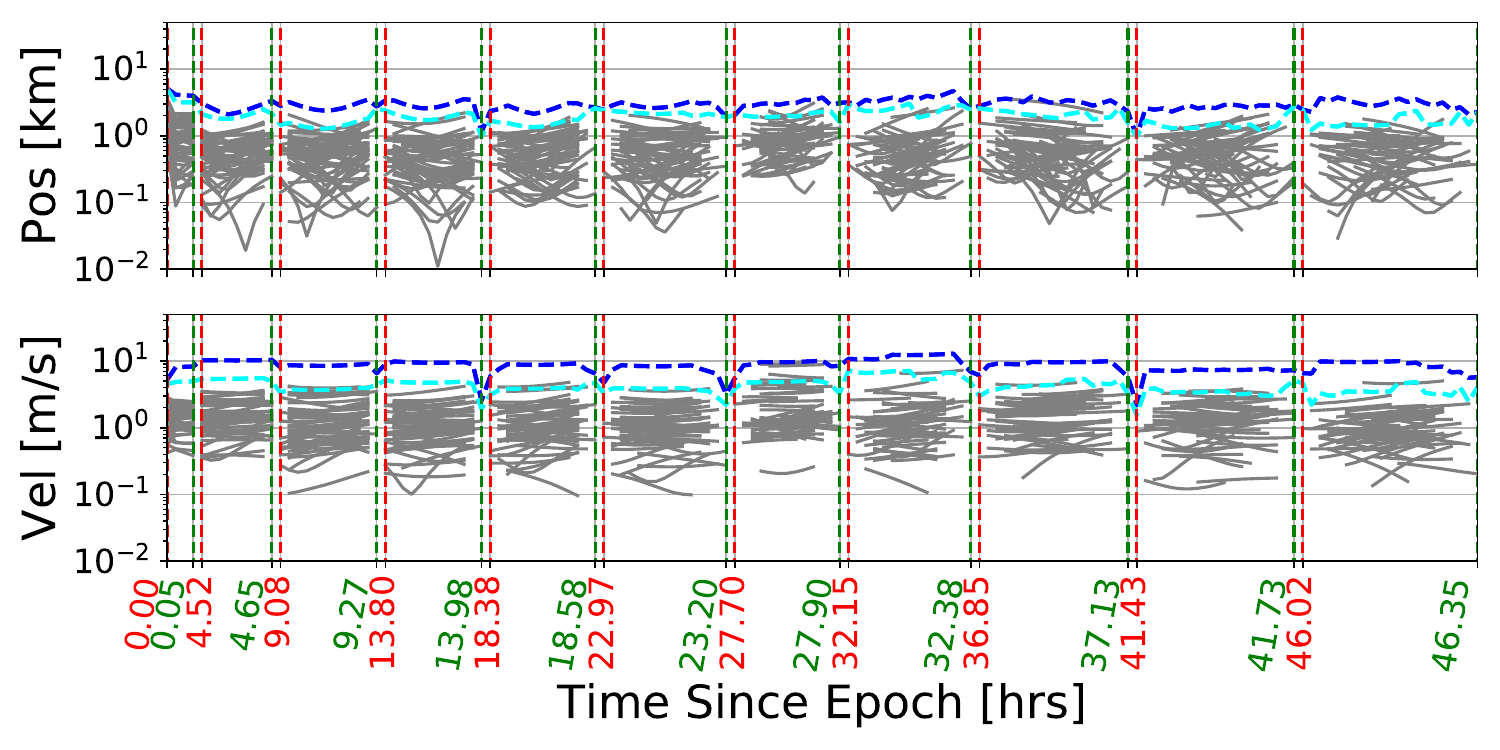}
    \caption{Estimate errors for the scenario of Sec.~\ref{sub:autocorr}.}
    \label{fig:autocorr} 
\end{figure*}

Since all previous scenarios only involve in-track only true thrust, either constant or intermittent, one scenario is included where the target accelerates with an almost continuously varying thrust in all three directions. The true thrust is drawn from a random process with constant segments of 60~seconds only. The standard deviation of the true thrust is 100~\textmu m/s\textsuperscript{2} in each direction, with an autocorrelation factor of 0.99. This means that portions of thrust profiles can be much larger than 300~\textmu m/s\textsuperscript{2}. The filter used is the same as Sec.~\ref{sub:manInit}. Figure~\ref{fig:autocorr} shows the results for a Monte Carlo of 50 trajectories. In every trajectory the filter can keep custody of the target, with few estimates falling outside of the 3$\sigma$ bounds. The RMSE is 763~m in position and 1.7~m/s in velocity.

\section{Conclusions}
\label{sec:conclusions}
This paper demonstrates the possibility of using Bayesian estimation to track low-thrust maneuvering spacecraft with sparse observations, while providing an explicit transitional prior. Having an explicit transitional prior is a strict requirement for implementation of any tracking technique into Bayesian multi-target tracking frameworks. The proposed method relies on multivariate Laplace distributions for the maneuvers, and focuses on the nonlinear transformation of the super-Gaussian tails of the distribution. This is achieved using principles from rare event simulation and through a novel $k$-nearest neighbor ensemble Gaussian mixture filter. Multi-fidelity methods are exploited to allow fast and efficient propagation. The resulting algorithm is robust to a variety of maneuver profiles and magnitudes, and is in large part insensitive to the tuning parameters. Further, the proposed approach does not require any additional maneuver detection techniques. The filter is able to maintain custody of targets that perform complex maneuvers involving intermittent thrust, even when the magnitude is up to 300~times larger than expected, and when measurement passes are sparse, one every 2.5~orbital revolutions or more. When observations are even sparser, fewer than two per day, the algorithm still maintains custody of targets with thrust magnitude up to 200~times larger than expected, at 200~\textmu m/s\textsuperscript{2}.

\section*{Funding Sources}

This material is based upon work supported by the Air Force Office of Scientific Research under award number FA9550-19-1-0404. Any opinions, findings, and conclusions or recommendations expressed in this material are those of the authors and do not necessarily reflect the views of the United States Air Force.

\end{document}